%% file: main.tex
\documentclass[10pt,journal,compsoc]{IEEEtran}
\ifCLASSOPTIONcompsoc
  \usepackage[nocompress]{cite}
\else
  \usepackage{cite}
\fi
\ifCLASSINFOpdf
\else
\fi

\usepackage{cite}

\usepackage{todonotes}
\usepackage{xspace}
\newcommand{\etal}{\textit{et al.}\xspace}

\newcommand{\ie}{\textit{i.e.,}\xspace}

\usepackage{subfigure}
\usepackage{titletoc}
\usepackage{setspace}
\usepackage{graphicx}
\usepackage{fancyhdr}
\usepackage{pdfpages}
\usepackage{setspace}
\usepackage{booktabs}
\usepackage{multirow}
\usepackage{caption}
\usepackage{tikz}
\usepackage{etoolbox}
\usepackage{hyperref}
\usepackage{xcolor}
\usepackage{color}
\usepackage{caption}
\usepackage{array}
\usepackage{amsmath, bm}
\usepackage{amssymb}
\usepackage{pdfpages}
\usepackage{listings}
\usepackage{float}
\usepackage{algorithm}  
\usepackage{algorithmicx}  
\usepackage{algpseudocode}
\usepackage{enumitem}
\usepackage{framed}
\usepackage{url}
\usepackage{cuted}
\usepackage{tcolorbox}
\usepackage{verbatim} 
\usepackage{amssymb}
\usepackage{makecell}

 \newtheorem{theorem}{Theorem}
  \newtheorem{proof}{Proof}

\begin{document}
%
\title{Zero-Knowledge Proof-based Practical Federated Learning on Blockchain}
%
%
%
%

\author{Zhibo~Xing,
        Zijian~Zhang$^{\ast}$,~\IEEEmembership{Member,~IEEE,}
        Meng~Li$^{\ast}$,~\IEEEmembership{Senior Member,~IEEE,}
        
        Jiamou Liu,
        Liehuang~Zhu,~\IEEEmembership{Senior Member,~IEEE,}
        Giovanni~Russello,~\IEEEmembership{Member,~IEEE,}
        and~Muhammad~Rizwan~Asghar,~\IEEEmembership{Member,~IEEE}
        
\IEEEcompsocitemizethanks{\IEEEcompsocthanksitem Zhibo Xing is with the School of Cyberspace Science and Technology, Beijing Institute of Technology, Beijing, China, and the School of Computer Science, The University of Auckland, Auckland, New Zealand. E-mail: 3120215670@bit.edu.cn.\protect\\

\IEEEcompsocthanksitem Zijian Zhang (Corresponding Author) is with the School of Cyberspace Science and Technology, Beijing Institute of Technology, Beijing, 100081, China, and Southeast Institute of Information Technology, Beijing Institute of Technology, Fujian, 351100, China. E-mail: zhangzijian@bit.edu.cn.\protect\\

\IEEEcompsocthanksitem Meng Li (Corresponding Author) is with Key Laboratory of Knowledge Engineering with Big Data (Hefei University of Technology), Ministry of Education; School of Computer Science and Information Engineering, Hefei University of Technology, 230601 Hefei, Anhui, China; Anhui Province Key Laboratory of Industry Safety and Emergency Technology; and Intelligent Interconnected Systems Laboratory of Anhui Province (Hefei University of Technology). Email: mengli@hfut.edu.cn.\protect\\

\IEEEcompsocthanksitem Jiamou Liu is with the School of Computer Science, The University of Auckland, Auckland, New Zealand. Email: jiamou.liu@auckland.ac.nz.\protect\\

\IEEEcompsocthanksitem Liehuang Zhu is with the School of Cyberspace Science and Technology, Beijing Institute of Technology, Beijing, 100081, China. E-mail: liehuangz@bit.edu.cn.\protect\\

\IEEEcompsocthanksitem Giovanni Russello is with the School of Computer Science, The University of Auckland, Auckland, New Zealand. E-mail: g.russello@auckland.ac.nz.\protect\\

\IEEEcompsocthanksitem Muhammad Rizwan Asghar is with the Department of Computer Science, University of Surrey, and the School of Computer Science, The University of Auckland, Auckland, New Zealand. E-mail: r.asghar@surrey.ac.uk.\protect\\

}
}

%
%

\markboth{Journal of \LaTeX\ Class Files,~Vol.~14, No.~8, August~2015}%
{Xing \MakeLowercase{\textit{et al.}}: Zero-Knowledge Proof-based Practical Federated Learning on Blockchain}
%



\IEEEtitleabstractindextext{%
\begin{abstract}
Since the concern of privacy leakage extremely discourages user participation in sharing data, federated learning has gradually become a promising technique for both academia and industry for achieving collaborative learning without leaking information about the local data.
Unfortunately, most federated learning solutions cannot efficiently verify the execution of each participant's local machine learning model and protect the privacy of user data, simultaneously.
In this article, we first propose a Zero-Knowledge Proof-based Federated Learning (ZKP-FL) scheme on blockchain. 
It leverages zero-knowledge proof for both the computation of local data and the aggregation of local model parameters, aiming to verify the computation process without requiring the plaintext of the local data. 
We further propose a Practical ZKP-FL (PZKP-FL) scheme to support fraction and non-linear operations. 
Specifically, we explore a Fraction-Integer mapping function, and use Taylor expansion to efficiently handle non-linear operations while maintaining the accuracy of the federated learning model.
We also analyze the security of PZKP-FL. 
Performance analysis demonstrates that the whole running time of the PZKP-FL scheme is approximately less than one minute in parallel execution.
\end{abstract}

\begin{IEEEkeywords}
Practical Zero-Knowledge Proof, Verifiable Federated Learning, zk-SNARK, Blockchain.
\end{IEEEkeywords}}

\maketitle

\IEEEdisplaynontitleabstractindextext

%
\IEEEpeerreviewmaketitle

\IEEEraisesectionheading{\section{Introduction\label{intro}}}

%
%
%
%
\input{introduction.tex}

\section{Related Works\label{related}}

\input{works.tex}

\section{Preliminaries\label{pre}}

\input{preliminaries.tex}

\section{Models and Goals\label{basic}}

\input{models.tex}

\section{The ZKP-FL Scheme\label{ZKP}}

\input{scheme.tex}

\section{The PZKP-FL Scheme\label{PZKP}}

\input{practical.tex}

\section{Security Analysis\label{Sanaly}}

\input{security.tex}

\section{Performance Analysis\label{Panaly}}

\input{performance.tex}

\section{Conclusions\label{con}}
In this work, we have proposed the ZKP-based Federated Learning (ZKP-FL) scheme to protect privacy while achieving accuracy in federated learning tasks. 
Specifically, we use ZKP and secure multi-party computation to prove the correctness of local computation and global model parameters without exposing local data and local model parameters. 
Then, we propose a Practical ZKP-FL (PZKP-FL) scheme to support fraction and non-linear operations by building a Fraction-Integer mapping and using the Taylor expansion. 
Both security and performance of the proposed schemes are also analyzed to demonstrate the PZKP-FL scheme is practical.


%



\ifCLASSOPTIONcompsoc
  \section*{Acknowledgments}
\else
  \section*{Acknowledgment}
\fi

This work is supported by National Natural Science Foundation of China (NSFC) under the grant No. 62172040,  No. 62002094, No. U1836212, and National Key Research and Development Program of China under the grant No.2021YFB2701200,  2022YFB2702402, and Anhui Provincial Natural Science Foundation under the grant No.2008085MF196.

\ifCLASSOPTIONcaptionsoff
  \newpage
\fi



%

\bibliographystyle{IEEEtran}
\bibliography{main}

%

\begin{IEEEbiography}[{\includegraphics[width=1in,height=1.25in,clip,keepaspectratio]{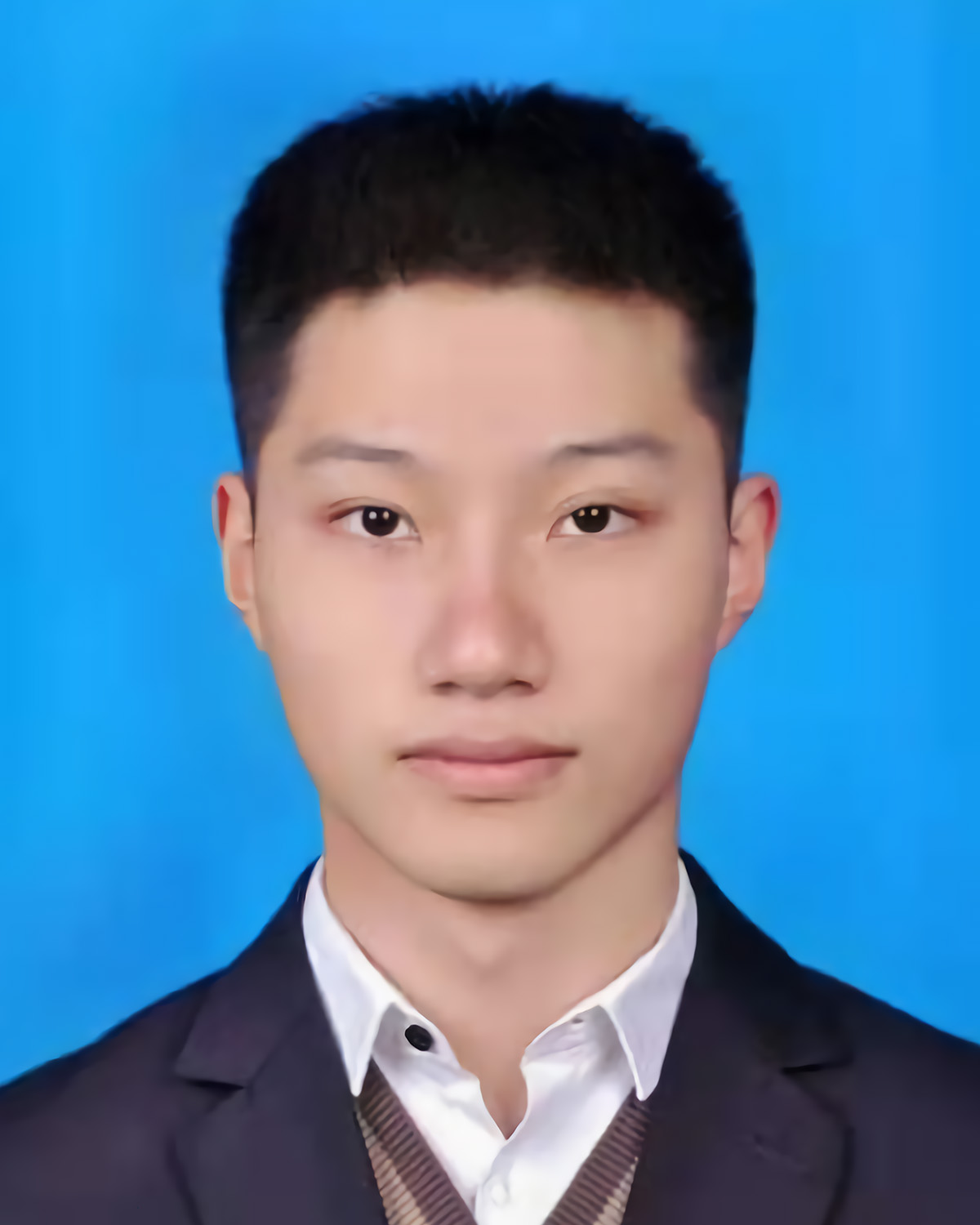}}]{Zhibo Xing} received the B.E. degree in computer science from Beijing Institute of Technology, Beijing, China, in 2017. He is currently pursuing the Ph.D. degree with the Cyberspace Science and Technology, Beijing Institute of Technology, Beijing, China, and the School of Computer Science, The University of Auckland, Auckland, New Zealand. His research interests include applied cryptography, data privacy and blockchain.
\end{IEEEbiography}
\begin{IEEEbiography}[{\includegraphics[width=1in,height=1.25in,clip,keepaspectratio]{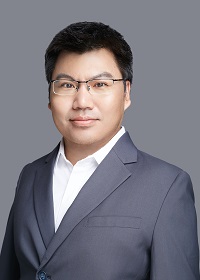}}]{Zijian Zhang}
received the Ph.D. degree from Beijing Institute of Technology, Beijing, China, in 2012. He is currently an Associate Professor with the School of Cyberspace Science and Technology, Beijing Institute of Technology. He is also a Research Fellow with the School of Computer Science, University of Auckland, Auckland, New Zealand. His research interests include data privacy and analysis of entity behavior and preference.
\end{IEEEbiography}
\begin{IEEEbiography}[{\includegraphics[width=1in,height=1.25in,clip,keepaspectratio]{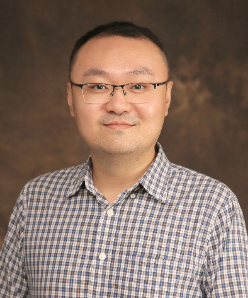}}]{Meng Li}
received the Ph.D. degree from Beijing Institute of Technology, Beijing, China, in 2019. He is currently an Dean Assistant with the School of Computer Science and Information Engineering, Hefei University of Technology, Hefei, China. He is also a Postdoc Research Fellow with the SPRITZ Security and Privacy Research Group, Department of Mathematics, University of Padua, Padua, Italy. His research interests include security, applied cryptography and cloud computing.
\end{IEEEbiography}
\begin{IEEEbiography}[{\includegraphics[width=1in,height=1.25in,clip,keepaspectratio]{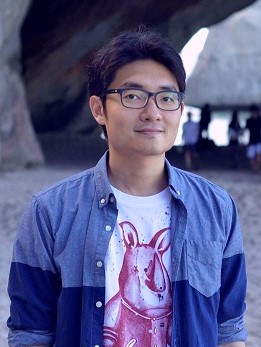}}]{Jiamou Liu}
received the Ph.D. degree from The University of Auckland, Auckland, New Zealand, in 2010. He is currently a Senior Lecturer with the School of Computer Science, The University of Auckland, Auckland, New Zealand. His research interests include artificial intelligence and data privacy.
\end{IEEEbiography}
\begin{IEEEbiography}[{\includegraphics[width=1in,height=1.25in,clip,keepaspectratio]{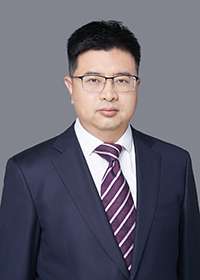}}]{Liehunag Zhu}
(Senior Member, IEEE) received the Ph.D. degree in computer science from Beijing Institute of Technology, Beijing, China, in 2004. He is currently a Professor and the Dean with the School of Cyberspace Science and Technology, Beijing Institute of Technology. His research interests
include security protocol analysis and design,
wireless sensor networks, and cloud computing.
\end{IEEEbiography}
\begin{IEEEbiography}[{\includegraphics[width=1in,height=1.25in,clip,keepaspectratio]{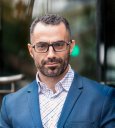}}]{Giovanni Russello}
received the Ph.D. degree from Eindhoven University of Technology (TU/e), Eindhoven, Netherland, in 2006. He is currently a Professor and the Head with the School of Computer Science, The University of Auckland, Auckland, New Zealand. His research interests include advanced access control, cloud computing security and applied cryptography.
\end{IEEEbiography}
\begin{IEEEbiography}[{\includegraphics[width=1in,height=1.25in,clip,keepaspectratio]{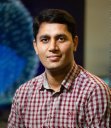}}]{Muhammad Rizwan Asghar}
received the Ph.D. degree from University of Trento, Trento, Italy, in 2013. He is currently a Reader with the Department of Computer Science, the University of Surrey, UK. He is also a Honorary Academic with the School of Computer Science, The University of Auckland, Auckland, New Zealand. His research interests include cybersecurity, applied cryptography and privacy.
\end{IEEEbiography}





\end{document}

%% file: introduction.tex
\IEEEPARstart{R}{ecently}, Artificial Intelligence (AI) has been widely used by both academic and industrial communities for many applications ranging from healthcare to commercial products.
For instance, Lampos \etal \cite{lampos2021artificial} explored a Machine Learning (ML) model for helping the communication with children with autism. 
Esteva \etal \cite{esteva2017dermatologist} trained a Convolutional Neural Network (CNN) by using a dataset of clinical images to classify skin cancer. 
AI is also a common tool in a variety of industrial products, including face recognition~\cite{he2005face} and personalized recommendation~\cite{shepitsen2008personalized}.
	
The existing AI learning process can generally be divided into two categories. 
One is centralized learning, where all of the training data is managed using a central server. %
This is the conventional way to train AI models and update model parameters. 
The other is decentralized learning, where the training data is distributed among multiple users.
\textcolor{black}{Each user has only parts of the training data as their local data.} 
In this case, each user firstly trains parameters with local data, and then integrates all of the local parameters into a global one. 
Users finally update their local model parameters by the global model parameters and iterate to train with their local data till the model converges. 
With the development of edge computing, the computation power of edge nodes extremely has grown significantly in recent years~\cite{shi2016edge}. 
Consequently, federated learning as advanced collaborative learning has become more widespread~\cite{mcmahan2017communication}.

Federated learning is aspired by enterprises for multiple reasons. 
Laws and regulations related to privacy protection expect enterprises to apply the federated training, such as the European Union's General Data Protection Regulation (GDPR)~\cite{voigt2017eu}, USA's California Consumer Privacy Act (CCPA)~\cite{CCPA}, and Personal Information Protection Law (PIPL) of the People's Republic of China~\cite{PIPL}. 
Once users' local data is leaked due to security vulnerabilities or illegally collected without their permission, then enterprises have to face a huge penalty. 
For an instance, Facebook was accused because it scanned faces in the user photo library and offered suggestions about who the person might be without user permission. 
Eventually, Facebook paid \$650 million to 1.6 million users~\cite{facebook}. 
Clearly, given the scale of information systems, massive data can be used for elaborating AI models. 
However, data are likely to be from multiple sources and requires multiple enterprises to collaborate. 
Since different enterprises have distinct requirements and goals, they might have some conflicts of interest.

Unfortunately, there is still a critical challenge in existing federated learning solutions. 
\textcolor{black}{More concretely, it is difficult to publicly verify all of the local training results without exposing local data.
Since local data are best to not be shared with all the participants for preventing the violation of data privacy, a malicious participant can generate local parameters by providing an appropriate random number rather than training with local data. }
Moreover, although local data are possessed by each user only, there are still several attacks based on the local parameters or gradients updated by user participants. 
For instance, Zhu \etal \cite{zhu2019deep} presented an optimized detection algorithm to obtain both local training input and label from the publicly shared gradients in a few iterations. 
Luo \etal \cite{luo2021feature} proposed a set of label inference attacks that achieved an outstanding performance to resist vertical federated learning. 
Lam \etal \cite{lam2021gradient} demonstrated that an unreliable central server can perform a dis-aggregate attack to recover the user participant matrix, which enabled traditional gradient inference attacks on users' personal training data. 
\textcolor{black}{Besides, Abadi \etal \cite{abadi2016deep} attempted to use differential privacy for effectively protecting users' local data. However, these works did not discussed the public verification for the local training process.}

There are existing works aiming at verification in federated learning.
Zhang \etal \cite{zhang2020zero} proposed a Zero-Knowledge Proof (ZKP) protocol for verifying the prediction result of a decision tree. 
Liu \etal \cite{liu2021zkcnn} introduced a ZKP scheme for verifying the prediction result of CNN. 
Weng \etal \cite{weng2021mystique} designed a ZKP system to prove the correctness of the training and utilization process.
However, these ZK protocols can only prove to one verifier at a time, and the communication cost is fairly high, when compared with succinct ZKPs like zk-SNARKs~\cite{groth2010short}. 
More importantly, the aforemnetioned works only supports the verification of the computational process in the use of the model, but not in the training process.  
Ruckel \etal \cite{ruckel2022fairness} investigated a federated learning system, enabling verification of the correctness of the training process. 
Nevertheless, their system only supports linear regression that is not a general approach for federated learning algorithms.


Considering the limitations of the existing solutions \cite{zhang2020zero, liu2021zkcnn, weng2021mystique, ruckel2022fairness}, 
\textcolor{black}{ZKP is intuitively used to achieve public verification and preserve data privacy in the training process for general AI models, simultaneously.
It can transform the training process into the corresponding arithmetic circuit constraints, so the verification can be achieved by proving the circuit satisfiability.}
\textcolor{black}{However, it brings a new challenge from the viewpoint of efficiency. }
\textcolor{black}{To the best of our knowledge, most existing zero-knowledge proof schemes are not practical to be directly applied for federated learning with scalable deep learning and machine learning models, 
because their computational cost is too heavy, and the privacy of intermediate parameters cannot be sufficiently guaranteed, which pose a privacy threat to the local data of each participant. }

\textcolor{black}{To solve all of the aforementioned problems together, in the local parameter generation process, we first separate a complex arithmetic circuit into multiple relatively simple sub-circuits. 
This is used to speed-up the verification process by parallel computing. Moreover, if the output of a sub-circuit is directly used as the input of the next sub-circuit, it can violate the privacy of local data. 
So, we construct a special ZK proof structure in each sub-circuit, and provide the corresponding ZK proof for the continuity of the input and output between two adjacent sub-circuits.
Furthermore, in the global parameter generation process, noise data are used for perturbing the original data, including the local gradients for deep learning model or local vectors for machine learning model. 
Meanwhile, we design a secure sum protocol on blockchain to achieve the public verification of global aggregation.
Besides, a practical FL scheme has to support the operations with float numbers and non-linear functions, because FL is involved in the fractional data types and various complex operations, such as exponential and logarithmic operations.
Consequently, we construct a special mapping algorithm to bridge the gap between float numbers and integers, such that an arbitrary machine learning algorithm can follow the arithmetic circuit constraints. }

The main contributions of this article are summarized as follows:
\begin{enumerate}
	\item A ZKP-based Federated Learning (ZKP-FL) scheme on blockchain is proposed to support the training AI models and provide formal security proof.
	\item A Practical ZKP-FL (PZKP-FL) scheme is further proposed to enable the ZKP-FL scheme to execute complex operations and handle fractions.
	\item \textcolor{black}{We provide both theoretical and experimental analyses, and the running time of the PZKP-FL scheme is approximately no more than one minute in the experiments.}
\end{enumerate}

The rest of this article is organized as follows. 
We review related work in Section \ref{related}. 
Next, Section \ref{pre} introduces some preliminaries.
After that, we explain the basic models and main idea in Section \ref{basic}.
Then, two new schemes ZKP-FL and PZKP-FL are proposed in Section \ref{ZKP} and Section \ref{PZKP}, respectively. 
Section \ref{Sanaly} provides security analysis. 
Section \ref{Panaly} reports performance analysis.
Finally, the conclusion is drawn in Section \ref{con}.

%% file: works.tex
This section describes the existing works in the past few years from three aspects, federated learning, privacy protection, and data confidentiality for federated learning. 
Specifically, federated learning is an advanced AI technique to achieve decentralized learning. 

The concept of federated learning was first introduced by Google \cite{mcmahan2017communication} in 2016 and has increasingly become popular topic in research. 
More related studies have been conducted in recent years. 
Hamer \etal \cite{hamer2020fedboost} proposed FedBoost, an ensemble learning approach for federated learning. 
It allows models limited by communication bandwidth or storage capacity could be trained by on-device data through federated learning. 
FedVision \cite{liu2020fedvision} is a ML platform to support the development of federated learning in computer vision applications. 
It helped customers to develop computer vision-based safety monitoring solutions in smart city applications and achieve efficiency improvements and cost reductions simultaneously. 
Liu \etal \cite{liu2020federated} presented a federated learning framework for vision-and-language grounding problems, including an aimNet network for converting both visual and textual features from image to a fine-grained representation. 
Blum \etal \cite{blum2021one} introduced a comprehensive game-theoretic framework for collaborative federated learning in the presence of agents who were interested in accomplishing their learning objectives while keeping their individual sample collection burden low.

There are several works to protect data confidentiality in federated learning. 
Considering differential privacy, FEDMD-NFDP \cite{sun2020federated} was a federated learning model in a distillation framework with a new noise-free differential privacy mechanism. 
This model guarantees each party's privacy without explicitly adding any noise, and can be proven to achieve ($\epsilon, \delta$)-differential privacy. 
LDP-FL \cite{sun2020ldp} was a local differential privacy mechanism for federated learning. 
It flexibly updated the local weights and adjusted the ranges of model parameters at different layers in a Deep Neural Network (DNN) by differential privacy. 
Kairouz \etal \cite{kairouz2021distributed} presented a comprehensive end-to-end system. 
By appropriate discretization of data and addition of discrete Gaussian noise, the system balances the dilemma between communication, privacy, and accuracy of aggregation. 
Wu \etal \cite{wu2022adaptive} proposed a federated learning scheme combined with the adaptive gradient descent strategy and differential privacy mechanism, which can protect the privacy of each computing participant from various background knowledge attacks with high stability of training and low communication cost. 
Li \etal \cite{li2020privacy} proposed a privacy-preserving federated learning framework based on an innovative chain-based Secure Multiparty Computation (SMC) technique. 
It can preserve privacy without lowering the model accuracy in the honest-but-curious setting with much lower communication and computation complexities than a typical SMC scheme. 
Based on ZKP, Asad \etal \cite{asad2021ceep} proposed a comprehensive approach that aims at reducing communication cost, preserving privacy on local gradients, and maintaining high accuracy.

Considering the verification of computation in machine learning, Zhang \etal \cite{zhang2020zero} introduced protocols for ZK decision tree predictions, allowing the owner of the model to convince others that the model computes a prediction on a data sample, without leaking any information about the model itself. 
Liu \etal \cite{liu2021zkcnn} investigated a ZKP scheme for CNN. 
The scheme allowed the owner of the CNN model to prove to others that the prediction of a data sample was indeed calculated by the model, without leaking any information about the model itself.  
Weng \etal \cite{weng2021mystique} designed a ZK system that allows proving that a submitted model is executed on the committed data or a committed model is executed on the submitted data. 
Zhao \etal \cite{zhao2021veriml} brought VeriML, a framework for integrity and fairness in outsourced machine learning, which support a total of six kind of models. 
Ruckel \etal \cite{ruckel2022fairness} also proposed a federated learning system that enabled users to validate that fellow clients indeed trained their submitted model updates based on the local data that they committed to.

%% file: preliminaries.tex
\subsection{zk-SNARKs}
zk-SNARK~\cite{groth2010short} is the zero knowledge succinct non-interactive arguments of knowledge.
zk-SNARK has two parts.
The former part, zk,~\cite{goldwasser1989knowledge} allows a prover to convince a verifier that a statement is true without revealing any other information, while the latter part, SNARK~\cite{groth2010short} provides a succinct proof of the correctness of circuit computations. 
Without loss of generality, ZKP has three properties, including completeness, soundness, and zero knowledge.

\emph{Groth16}~\cite{groth2016size} is a pairing-based zk-SNARK for arithmetic circuit satisfiability. 
In brief, it has four functions $(\mathrm{Setup, Prove, Vfy, Sim})$: 
$(\sigma,\tau)\leftarrow\mathrm{Setup}(R):$ The setup take a security parameter $\lambda$ and a relation $R\in R_\lambda$ as input, outputs a common reference string $\sigma$ and a simulation trapdoor
$\tau$ for the relation $R$.
$\pi\leftarrow\mathrm{Prove}(R,\sigma,\phi,w):$ The prove takesa common reference string $\sigma$ and $(\phi,w)\in R$ as input, outputs argument $\pi$.
$0/1\leftarrow\mathrm{Vfy}(R,\sigma,\phi,\pi):$ The vfy takes a common reference string $\sigma$, a statement $\phi$ and an argument $\pi$ as input, outputs 0 (reject) or
1 (accept).
$\pi\leftarrow\mathrm{Sim}(R,\tau,\phi):$ The sim takes a simulation trapdoor $\tau$ and statement $\phi$ as input, outputs an argument $\pi$.

\subsection{$\Sigma$-Protocol}
A $\Sigma$-protocol~\cite{damgaard2002sigma} proves discrete logarithm relation without revealing the witness. 
It constructs a cryptographic primitive for ZKF. 
Without loss of generality, a $\Sigma$-protocol has three properties, completeness, special soundness, and honest verifier ZK.

\subsection{Secure Sum Protocol} 
\label{ssp}
Bonawitz \etal \cite{bonawitz2017practical} proposed a one-time pad method to compute the sum $X=\Sigma_{i=1}^n x_i$ without revealing each party's private factor $x_i$ to any other parties. 
Several modifications have been made to the original scheme. 
In this protocol, there are $n$ user nodes and one central node.
Each user node has a secret value $x_i$ to be aggregated, the center has a pair of keys $pk$, $vk$ for additive homomorphic encryption, and the public key has been sent to all users. 
The protocol specifications are as follows.

\begin{enumerate}
	\item First, for each pair of user nodes $U_i$ and $U_{i+1}$ ($U_n$ can be paired with $U_1$), exchange a secret value $s_i$.
	
	\item For each user node $u_i$, $u_i$ computes $x'_i=x_i+s_i-s_{i-1}$ (we define $s_0=s_n$), then encrypt $x'_i$ with $pk$ as $c_i\leftarrow Enc_{pk}(x'_i)$. $u_i$ sends $c_i$ to the central node.
	
	\item As the central node has collected all the $c_i$, it computes the final result as $\Sigma_{i=1}^n x_i=Dec_{sk}(\Sigma_{i=1}^n c_i)$.
	
\end{enumerate}

Considering the additive homomorphic encryption scheme, each $s_i$ is eliminated.

\subsection{Paillier Encryption Scheme}
\label{pai}
Paillier \cite{zhang2020zero} is an additive homomorphic encryption scheme, which consists of four algorithms, including KeyGen, Enc, Dec and Add.

\begin{itemize}
	\item $KeyGen(n)\rightarrow (pk, sk)$: Randomly choose two primes $p,q$, which meet $gcd(pq,(p-1)(q-1))=1$. 
	Compute $n=pq, \lambda=lcm(p-1,q-1)$. 
	Randomly select $g\in \mathbb{Z}_{n^2}^{*}$. 
	Set $pk=<n,g>, sk=<\lambda>$.
	
	\item $Enc_{pk}(m)\rightarrow c$: Randomly choose $r\in\mathbb{Z}_n^*$. 
	Compute $c=g^mr^n \rm{\ mod\ } n^2$.
	
	\item $Dec_{sk}(c)\rightarrow m$: Set $L(x)=\frac{x-1}{n}$, then compute $m=\frac{L(c^\lambda \rm{\ mod\ } n^2)}{L(g^\lambda \rm{\ mod\ } n^2)} \rm{\ mod\ } n$.
	
	\item $Add(c_x, c_y)\rightarrow c_z$: 
	Considering $c_x = Enc_{pk}(x)$ and $c_y = Enc_{pk}(y)$, then $c_z$ satisfies that $Dec_{sk}(c_z)=x+y$.

\end{itemize}

\subsection{Blockchain and Smart Contract} 
\textcolor{black}{Blockchain~\cite{nakamoto2008bitcoin} was first proposed as the underlying key technology of Bitcoin, which enables transactions between unfamiliar nodes without relying on trusted third parties availiable. And smart contract~\cite{szabo1997formalizing}, an earlier proposed technology, are better empowered by the presence of blockchain. The original intent of smart contract was to digitize a set of commitments to enable the various participants to follow the committed steps of the protocol. This makes it difficult for the participants to do evil and the protocol can achieve the desired results. The trusted execution and decentralized nature of blockchain for transactions makes smart contracts no longer remain a concept, but actually a program that can run on the blockchain system to execute transactions automatically.}

%% file: models.tex
\subsection{System Model}
\textcolor{black}{Traditional federated learning involves two kinds of participants, a publisher and several trainers.
The publisher first issues an AI model with initiative parameters.
The trainers run the AI model based on their local data, and they report the local parameters. 
The publisher keep aggregating all of the local parameters and starting a new round with the aggregated parameters till the AI model converges.
For the decentralized FL, all of the trainers can negotiate to decide the AI model with the initiative parameters and the aggregation rules at the start.
They can use any decentralized techniques like blockchain to be the communication channel for exchanging the local and aggregated parameters.}

\textcolor{black}{In order to achieve the verification of the training process and ensure the privacy of the local data at the same time, the system model in this paper is slightly adjusted as below.}
First, the publisher issues an AI model with initiative parameters as the training task.
Then, all the trainers run the AI model with their local data and generate ZKPs for the training process. 
Here, ZKPs are used to convince the publisher that the local model is correctly computed. 
\textcolor{black}{Following by that, the publisher computes the aggregated parameters by executing a secure sum protocol via a well-designed smart contract on blockchain. 
The secure sum protocol essentially plays the role of the aggregation rule.
And the smart contract guarantees trusted parameter aggregation result. 
All the above steps are continuously executed till the AI model converges.
The adjustment is considered for protecting local data from being exposed by other trainers and the publisher from analyzing the intermediate parameters, such as the local and aggregated gradients of an artificial neural networks. }

\begin{figure*}[htbp]
	\centering
	\includegraphics[width=6.5in]{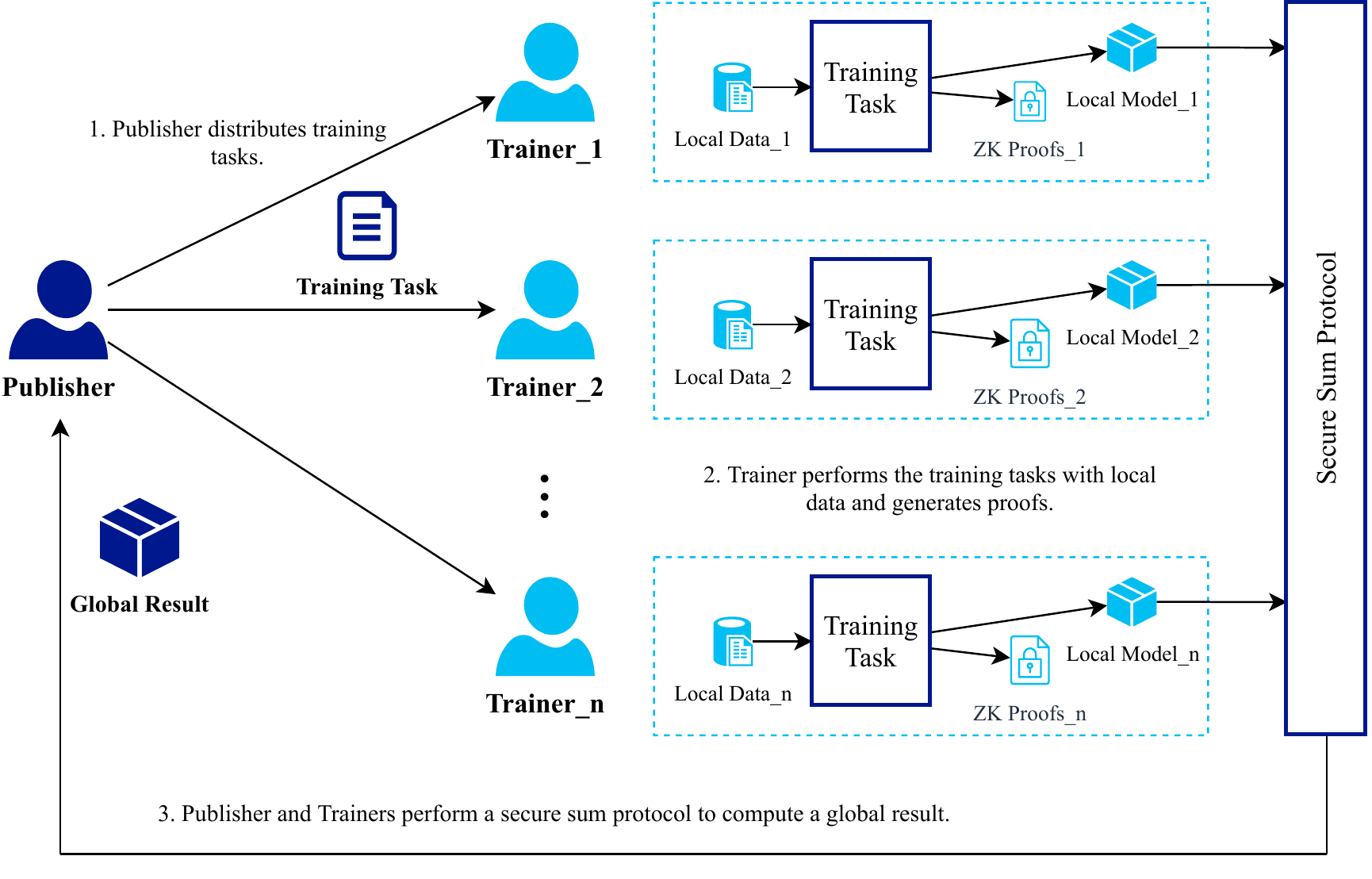}
	\caption{The process of the ZKP-FL scheme.}
	\label{framework}
\end{figure*}

\subsection{Threat Model}
\label{sec:threat}
In this work, two kinds of adversaries are considered, including the lazy but curious trainer and the curious and unreliable publisher. 
The lazy participants will attempt to do computational works as less as possible. 
The curious participants will inspect others' local data from the available information. 
\textcolor{black}{The unreliable participants may provide incorrect information or perform different computations than expected. 
That is to say, the lazy but curious trainer tries to: 
\begin{itemize}
	\item cheat the server by giving parameters generated in some other ways rather than training, so that the trainer can get rewards with less workload.
	\item steal other trainers' local data from the local parameters they trained and submitted.
\end{itemize}
}

The reason we define the trainers as lazy but curious is that we assume the trainers are rational. 
This means they will not take malicious actions without gain. 

\textcolor{black}{The curious and unreliable publisher tries to:
\begin{itemize}
	\item trick the user into accepting an incorrect aggregation of global parameter due to the adversarial behaviors in multi-client verifiable computation~\cite{choi2013multi}.
	\item steal trainers' local data from the local parameters they trained and submitted.
\end{itemize}
}
\subsection{Goals}
\textcolor{black}{In the FL scenarios, we set three goals from the viewpoint of security.}

\textbf{The Public Verification of Local Computation}

\textcolor{black}{The publisher and all of the trainers must enable to verify whether all of the local parameters are computed by using some local data.
In other words, the local parameters are difficult to be generated by simply selecting from random numbers.}

\textbf{The Public Verification of Global Aggregation}

\textcolor{black}{All of the trainers must enable to verify whether the aggregated parameters was correctly executed by using the secure sum protocol.}

\textbf{The Privacy Protection of Local Data}

\textcolor{black}{The FL does not disclose additional information about the local data to any other participants in the process. 
That is, the FL schemes have to protect participants from both kinds of adversaries outlined in Section \ref{sec:threat}, by leveraging the ZKP and secure sum protocol.}

%% file: scheme.tex
In this section, we first introduce the overall process of the ZKP-FL scheme step-by-step.
The whole procedure is divided into 3 phases: model distribution, model training, and model aggregation. 

\begin{enumerate}
    \item In the model distribution phase, the publisher processes and distributes the training tasks and models.
    \begin{itemize}
        \item The publisher divides the training algorithm $F$ into $q$ identical pieces $P$.
        
        \item The publisher converts the algorithm piece $P$ into arithmetic circuit constraints $R$ for $Groth16$, then runs the $Groth16.Setup()$ algorithm to generate the common reference string $(\sigma, \tau)$, containing the proving key $pk$ and verification key $vk$, for both ZKP generation and verification.
        
        \item The publisher sends the algorithm piece $P$, corresponding CRS $(\sigma, \tau)$, and constraints $R$ to $n$ trainers.
        
    \end{itemize}
    
    \item In the model training phase, each trainer runs the training algorithm to train the local model, and generates the corresponding ZKPs for the correctness of the training process.
    \begin{itemize}
        \item Trainer $i$ trains the model using local data $d_i$ by running the algorithm piece $P$ for several rounds, and keeps all the inputs and outputs in each round as statements $\phi_i$.
        
        \item Trainer $i$ generates the ZKPs $\pi_i$ of each round of the training process by running the $Groth16.Prove()$ algorithm. 
        To avoid the exposure of intermediate inputs and outputs of the training process to the publisher during the proof verification, Trainer $i$ runs both Algorithm \ref{modify} and Algorithm \ref{prove modification}. 
        The former algorithm is the $Groth16.Prove()$ algorithm with some extra modification, which aims at a slight modification on statement $\phi_i$ and the verification key $vk_i$ into $\phi'_i$ and $vk'_i$, respectively, to conceal the data. 
        The latter one is for generating ZKP $s^1_i$ and $s^2_i$ arguing that the modification on the statement and verification key is valid, and the output of the previous piece is the input to the next piece.
        
        \item After all the training and proof generation, Trainer $i$ sends all the proofs and modified data to the publisher. 
        The modified local model is contained in the last modified statement as the final output.
    \end{itemize}

    \item In the model aggregation phase, the publisher and $m$ trainers whose proofs are verified run a secure sum protocol to compute the global model without revealing any local models.
    \begin{itemize}
        \item Upon receiving Trainer $i$'s proofs $\pi_i,s^1_i,s^2_i$ and modified data $\phi'_i,vk'_i$, the publisher runs the $Groth16.Verify()$ algorithm to verify proof $\pi_i$ and runs Algorithm \ref{verify modification} to verify proof $s^1_i$ and $s^2_i$, aiming at ensuring that Trainer $i$'s training process is carried out correctly.
        
        \item The publisher runs the $KenGen()$ algorithm to get the public key $pk_h$ and secret key $sk_h$ for homomorphic encryption, and chooses a generator $g_{pub}\in G$ randomly. 
        Then, it sends $pk$ and $g_{pub}$ to trainers whose proof is verified.
        \textcolor{black}{Also, it sets up and exposes a smart contract $SC$ for the final computation of the parameter aggregation. 
    }
        \item Trainer $i$ generates and sends random values $s_{i,i+1}\in Z_q$ to Trainer $i+1$. 
        Note that the last trainer $m$ sends to the first trainer $1$.
        
        \item Trainer $i$ computes $c_{i}=Enc_{pk}(\phi_{i}+s_{i}-s_{i-1})$ (we define $s_{0}=s_{m}$). \
        Trainer $i$ also runs Algorithm \ref{prove sum} to obtain proof $s^3_i$, arguing that $c_{i}$ contains the same output as $\phi'_{i}$.
        
        \item \textcolor{black}{Trainer $i$ sends $c_i$ and $s^3_i$ to the smart contract $SC$. }
        
        \item Upon receiving $c_i$ and $s^3$ from Trainer $i$, the publisher computes the global model $\bar{c}=\frac{Dec_{sk}(\Sigma_{i=1}^m c_{i})}{n}$ through the smart contract $SC$. 
        Then, the publisher runs Algorithm \ref{verify sum} to check the correctness of the submitted models involved in the aggregation of the global model.
    \end{itemize}
\end{enumerate}

The whole process of the ZKP-FL scheme is shown in Fig.~\ref{framework}.

\subsection{Model Distribution}
Intuitively, as long as we can convert the training algorithm into an arithmetic circuit, we can generate the ZKP arguing for the correctness of the training process. 
However, there could be memory limitations as the typical training algorithm might be too big to be converted into a circuit. 
For a simple gradient descent algorithm, consider 90 input samples and iterate through 100 rounds, more than one hundred gigabytes of memory is required for converting the entire computation process into a circuit. 
Hence, the trainer has to split the original algorithm into pieces to make the conversion possible and practical. 
Then, the trainer can generate several smaller circuits for ZKPs instead of the big ones. 
The division of the training algorithm not only decreases the size of the algorithm, but also improves the efficiency, because the conversion is smaller and only needs to be done once. 
Besides, all the proofs can be generated in parallel after the training.

In most ML algorithms, the training process includes multi-round iterative computations. 
Let $p'=F(d, p)$, where $F$ denotes the ML training algorithm, $d$ denotes the local data, $p$ denotes the original model parameter and $p'$ denotes the well-trained parameters.
Suppose that in this process some computational procedure $A$ is executed $N$ times by the loop, i.e. $F(d,p):{\rm for}(i=1..N)A(d,p)$.
We divide the ML algorithm $F$ into $q$ identical pieces $P$, labeled $1$ to $q$. 
$q$ should be an integer factor of $N$, then $P$ is a composition of one or several $A$. 
By default, we set $q = N$, \ie treat each iteration as a piece.
Due to the division method, each piece of algorithm is the same. 
Using the input data $d$, output parameters $p$, and the circuit transformed from the algorithm, we can generate a ZKP for the correctness of the computation process without revealing the private input data. 

\begin{figure*}
	\centering
	\includegraphics[width=6.5in]{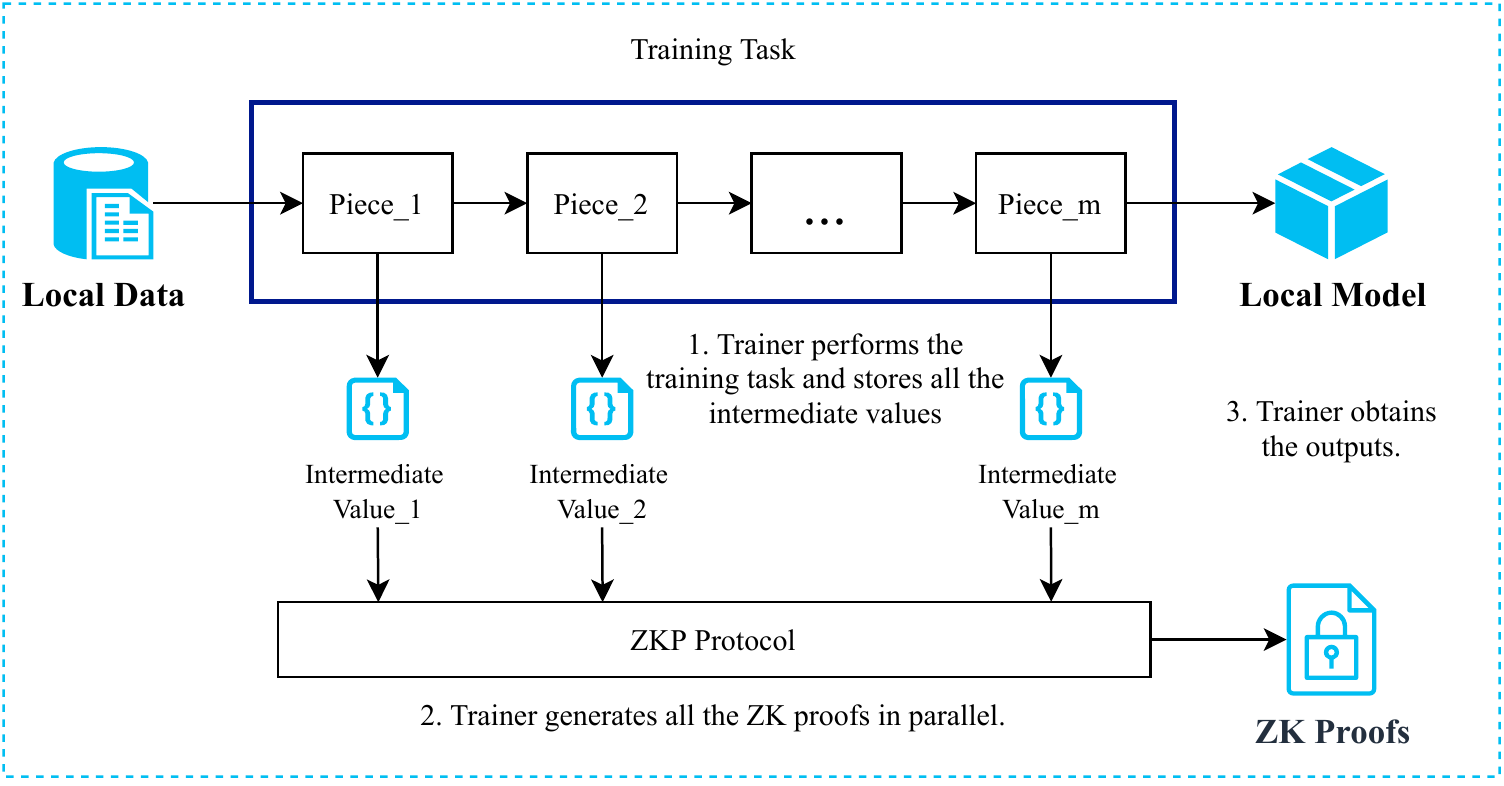}
	\caption{The training process of the ZKP-FL scheme.}
	\label{breakdown}
\end{figure*}

We use $Groth16$ as the ZKP scheme, which requires a trusted setup. 
As no trainer is considered trustworthy, the publisher has to run the setup and send all trainers both the training algorithm and the corresponding generated Common Reference String (CRS), including the proving key $pk$ and the verification key $vk$, describing the arithmetic circuit. 
As the model distribution phase is completed, all the trainers get the piece of training algorithm for training and corresponding CRS for generating ZKPs.

\subsection{Model Training}
In this subsection, we ignore for now the use of $i$ as the label of trainer and instead, we use $i$ for the piece.

The verification algorithm for $Groth16$ is $Verify(R, \sigma, \phi, \pi) \rightarrow \{0, 1\}$, which outputs 1 when equation \ref{groth16verify} is satisfied, where $\phi$ denotes the explicit inputs and outputs of the circuit, $\sigma$ denotes the CRS containing the verification key $vk$ and proving key $pk$. 
\begin{align}
    \label{groth16verify}
    e(A,B)=e(G^\alpha,H^\beta)&\cdot
    e(G^{\frac{\Sigma_{i=0}^la_i(\beta u_i(x)+\alpha v_i(x)+w_i(x))}{\gamma}}, H^\gamma)\notag\\
    &\cdot e(C,H^\delta)\notag\\
    =e(G^\alpha,H^\beta)&\cdot
    e(\Pi_{i=0}^l (G^{\frac{(\beta u_i(x)+\alpha v_i(x)+w_i(x))}{\gamma}})^{a_i}, H^\gamma)\notag\\
    &\cdot e(C,H^\delta)
\end{align}
Some modifications have been made to the original $Groth16$ so that it can avoid the exposure of both inputs (private training data) and outputs (intermediate local model) to the publisher for verification. 
Considering the attack described in \cite{zhu2019deep}, the leakage of the intermediate gradient is dangerous.
We add uniformly distributed noise $t$ to statement $\phi$ for protecting. 
To balance the impact of $t$ for the verification process, a modification is implemented on the verification key $vk$, which enables the publisher to verify the proof with the original verification algorithm. 
In particular, we define the inputs to the algorithm circuit to be the training data and the model parameters, and the outputs to be the training data and the optimized model parameters.
For $i$ from $1$ to $l/2$, we consider $\phi_i$ as the input model parameter; while, for $i$ from $l/2+1$ to $l$, we consider $\phi_i$ as the output optimized parameter. 
As for the local data, we place them in the implicit witness $w$. 
The training process is shown in Fig. \ref{breakdown}.

Algorithm \ref{modify} is the modified $Groth16$ proof generation algorithm. 
Algorithm \ref{prove modification} is for generating the proof, arguing that the modification on the verification key and statement made by Algorithm \ref{modify} is valid. 
Algorithm \ref{verify modification} is for verifying the proof generated in Algorithm \ref{prove modification}.

\begin{algorithm}
	\caption{{\rm \textbf{Gen\_G16\_Prf (.)}}: Proof generation of the modified $Groth16$ for Piece $i$. \label{modify}}
	\hspace*{0.07in}{\bf Input:}
	Proving key $pk$, verification key $vk$, statement $\phi_i=\{a_{i,j}\}_{j=1}^l$, witness $w_i=\{a_{i,j}\}_{j=l+1}^m$, t-list of piece $i-1$ $T_{i-1}=\{t_{i-1,j}\}_{j=1}^l$.\\
	\hspace*{0.07in}{\bf Output:}
	The modified verification key $vk'_i$, modified statement $\phi'_i$, ZKP $\pi_i$, t-list of Piece $i$, $T_i$, checker of Piece $i$ $tsum^1_i, tsum^2_i$.
	\begin{algorithmic}[1]
		\State Run algorithm $Prove()$ in Groth16 with proving key $pk$, statement $\phi_i$, witness $w_i$ to get ZKP $\pi_i$
		\State Set $vk'_i=vk$, $tsum^1_i=0$, $tsum^2_i=0$
		\For{$j$ in $1 \ldots l$}
		\If{$j\leq l/2$}
		\State $t_{i,j}=t_{i-1, j+l/2}$
		\State $tsum^1_i*=vk_i.\gamma\_abc_j^{-t_{i,j}}$
		\ElsIf{$k\geq l/2+1$}
		\State Randomly choose $t \in_R Z_q$, $t_{i,j}=t$
		\State $tsum^2_i*=vk_i.\gamma\_abc_j^{-t_{i,j}}$
		\EndIf
		\State Compute $a'_{i,j}=a_{i,j}+t$
		\State Compute $vk_i'.\gamma\_abc_0 = vk_i'.\gamma\_abc_0 \cdot vk_i.\gamma\_abc_j^{-t_{i,j}}$
		\EndFor
		\\
		\Return $vk_i'$, $\phi_i'$, $tsum^1_i$, $tsum^2_i$, $\pi_i$, $T_i$
	\end{algorithmic}
\end{algorithm}

Note that $T_0$ is defined as $\left\{t_{0,j} | t_{0,j}\in_R Z_q, j=1, \ldots ,l\right\}$, $vk_i.\gamma\_abc_j$ stands for $G^\frac{\Sigma_{j=0}^l(\beta u_j(x)+\alpha v_j(x)+w_j(x))}{\gamma}$ in $Groth16$. 

For the completeness of the modification, the verification process with modified $vk'$ and $\phi'$ is shown in equation \ref{modifiedverify}, where we ignore the piece $i$ for the sake of clarity of the equation.
\begin{align}
    \label{modifiedverify}
    \Pi_{j=0}^l (vk'.\gamma\_abc_j)^{a_j'}
    & = \Pi_{j=1}^l (vk.\gamma\_abc_j)^{a_j+t_j}\notag\\
    & \quad \cdot vk.\gamma\_abc_0\cdot \Pi_{j=1}^l vk.\gamma\_abc_j^{-t_j}\notag\\
    & = \Pi_{j=1}^l (vk.\gamma\_abc_j)^{a_j}\cdot vk.\gamma\_abc_0^{a_0}\notag\\
    & \quad \cdot \Pi_{j=1}^l (vk.\gamma\_abc_j)^{t_j} \cdot \Pi_{j=1}^l vk.\gamma\_abc_j^{-t_j}\notag\\
    & = \Pi_{j=0}^l (vk.\gamma\_abc_j)^{a_j}
\end{align}
In equation \ref{modifiedverify}, the first equal sign depends on lines 11 and 12 of Algorithm \ref{modify}, and the second equal sign depends on $a_0=1$.
Therefore equation \ref{modifiedverify} holds, and thus the verification process for the modified $vk'$ and $\phi'$ is the same as that before the modification, so this modification does not affect the correctness of the verification process.

For the soundness of the modification, the prover has to prove that $vk_i'$ is indeed a kind of modification of $vk_i$, and the prover has the knowledge of that modification. 
Further, the prover also has to prove the consistency between the output $a_{i-1,l/2+j}$ in piece $i-1$ and the input $a_{i,j}$ in piece $i$.
So, we consider a $\Sigma$-protocol to address this aspect. 
That is, we guarantee its soundness by Algorithm \ref{prove modification} and Algorithm \ref{verify modification}.

Since $t$ is a random number, the adversary can concatenate multiple inputs and outputs using different $t$. 
To this end, we add an extra $\Sigma$-protocol for checking the consistency to ensure the same $t$ is used. 
Furthermore, to ensure that the previous piece's output is the following piece's input, a $\Sigma$-protocol proof is needed. 
The $\Sigma$-protocol proof generation and verification algorithm is shown in Algorithms \ref{prove modification} and \ref{verify modification}.

\begin{algorithm}
	\caption{{\rm \textbf{Gen\_Con\_Prf(.)}}: Proof generation to combine the pieces \label{prove modification}}
	\hspace*{0.07in}{\bf Input:}
	The verification key $vk$, the modified verification key $vk_i'$, t-list $T_i=\{t_{i,j}\}_{j=1}^l$, checker $tsum^1_i, tsum^2_{i-1}$.\\
	\hspace*{0.07in}{\bf Output:}
	The sigma proof $s^1_i$, $s^2_i$.
	\begin{algorithmic}[1]
		\For{$j$ in $1 \ldots l$}
		\State Set $c_j=t_{i,j}, g_j=vk_i.\gamma\_abc_j$
		\EndFor
		\State Set $C_1=tsum^1_i, C_2=tsum^2_{i-1}, C=vk_i'.\gamma\_abc_0\cdot vk_i.\gamma\_abc_0^{-1}$ 
		\State Run $\Sigma$-protocol to obtain proof $s^1_i=PoK\{(c_1, \ldots,c_l,$
		$r_1, \ldots,r_l):C=\Pi_{i=1}^lg_i^{c_i}\}$
		\If{$i \neq 1$}
		\State Run $\Sigma$-protocol to obtain proof $s^2_i=PoK\{(c_1, \ldots,$
		$c_{l/2},r_1, \ldots,r_{l/2}):C_1 = \Pi_{i=1}^{l/2}g_{i}^{c_i}, C_2 = \Pi_{i=1}^{l/2}g_{i+l/2}^{c_i}\}$
		\Else
		\State Set $s^2_i=null$
		\EndIf
		\\
		\Return $s^1_i, s^2_i$
	\end{algorithmic}
\end{algorithm}

\begin{algorithm}
	\caption{{\rm \textbf{Vrf\_Con\_Prf(.)}}: Proof verification of the combined pieces\label{verify modification}}
	\hspace*{0.07in}{\bf Input:}
	The sigma proof $s^1_i, s^2_i$, the verification key $vk_i$, the modified verification key $vk_i'$, checker of Piece $i-1$, $tsum^1_{i-1}$, checker of Piece $i$, $tsum^1_i, tsum^2_i$.\\
	\hspace*{0.07in}{\bf Output:}
	The verification result $b$.
	\begin{algorithmic}[1]
		\If{$vk_i.\gamma\_abc_0 \cdot tsum^1_i \cdot tsum^2_i == vk_i'.\gamma\_abc_0$}
		\For{$j$ in $1 \ldots l$}
		\State Set $g_j=vk.\gamma\_abc_j$
		\EndFor
		\State Set $C=vk'.\gamma\_abc_0\cdot vk.\gamma\_abc_0^{-1}$
		\If{the verification of $s^1_i$ is successful}
		\State Set $C_1=tsum^1_i, C_2=tsum^2_{i-1}$
		\If{the verification of $s^2_i$ is successful}
		\State \textbf{return} true
		\EndIf
		\EndIf
		\EndIf
		\\
		\Return false
	\end{algorithmic}
\end{algorithm}

The completeness and soundness of $s^1$ and $s^2$ simply relies on the $\Sigma$-protocol. 
Thus $s^1_i$ and $s^2_i$ guarantee that the equivalence of each noise $t_{i,j}$ used for the input on $a_{i,j}$ in round $i$ and the noise $t_{i-1,l/2+j}$ used for the output on $a_{i-1,l/2+j}$ in round $i-1$.
This ensures not only the correctness of the added noise, which is the correctness of $vk'$ and $\phi'$ as a modified version of $vk$ and $\phi$.
But also due to $a_{i,j}'$ is explicit for the verifier, if $s^1$ and $s^2$ pass the verification then the verifier can believe that the corresponding input and output $a_{i,j}$ used in the two rounds is the same based on lines 11 of Algorithm \ref{modify}.
In this way the soundness of the modification is guaranteed, for the noise $t$ is strictly limited and verified.

Specifically, the $\Sigma$-protocol for $s^1$ is to check if the equation \ref{s1sigma} holds, where $e=Hash(g_1||...||g_l||C||g_{1}^{r_1}||...||g_{l}^{r_l})$.
\begin{align}
    \label{s1sigma}
    \Pi_{i=1}^l g_i^{r_i+ec_i} \equiv \Pi_{i=1}^l g_i^{r_i}\cdot C^{e} \mod p
\end{align}
And the $\Sigma$-protocol for $s^2$ is to check if the equation \ref{s2sigma1} and equation \ref{s2sigma2} hold, where $e=Hash(g_1||...||g_{l/2}||C_1||C_2||g_{1}^{r_1}||...||g_{l/2}^{r_{l/2}})$.
\begin{align}
    \label{s2sigma1}
    \Pi_{i=1}^{l/2} g_i^{r_i+ec_i} \equiv \Pi_{i=1}^{l/2} g_i^{r_i}\cdot C_1^{e} \mod p\\
    \label{s2sigma2}
    \Pi_{i=1}^{l/2} g_{i+l/2}^{r_i+ec_i} \equiv \Pi_{i=1}^{l/2} g_{i+l/2}^{r_i}\cdot C_2^{e} \mod p
\end{align}

As the model training phase is completed, the trainer sends all the modified statements $\phi'$, the modified verification key $vk'$, and the $\Sigma$ proof $s^1$ and $s^2$ to the publisher.

\subsection{Model Aggregation}
The modified statement $\phi'$ contains the noised local model, which cannot be extracted by the publisher directly. 
So, by running a multiparty secure sum protocol with several trainers, the publisher can compute a global model without revealing any local model to any party. 
The secure sum protocol is described in Section \ref{ssp} and the cryptosystem is described in Section \ref{pai}. 
The specification of this secure sum protocol is described as follows:

\begin{enumerate}
	\item \textbf{Initialization.}\\
	The publisher runs $KeyGen(n)$ to get the public key $pk$, the secret key $sk$, chooses a generator $g_{pub}$ in group $G$, and sends $pk$ and $g_{pub}$ to all trainers. 
	The publisher also sets up and exposes a smart contract $SC$ to all trainers for the computation of the global model. 
	Trainer $i$ generates and sends random values $s_{i,j}\in Z_q$ where $j=1, \ldots,l-1$ to the trainer $i+1$. 
	Note that the last trainer $l$ sends them to the first trainer.
	
	\item \textbf{Secure Sum Computation.}\\
	Trainer $i$ computes $c_{i,j}=Enc_{pk}(a_{i,j}+s_{i,j}-s_{i-1,j})$ for $j=1, \ldots,l$ (we define $s_{0,j}=s_{n,j}$). 
	Trainer $i$ also runs Algorithm \ref{prove sum}($\phi_i,T_i,g_{pub}$) to obtain proof $S^3_i$ and corresponding generators $g_{i,1}, \ldots, g_{i,l}$ and commitments $C_{i,1}, \ldots,C_{i,l}$. 
	\textcolor{black}{Each Trainer $i$ sends $c_{i,1}, \ldots, c_{i,l}, S^3_i, g_{i,1}, \ldots, g_{i,l}, C_{i,1}, \ldots, C_{i,l}$ to the smart contract $SC$. 
	The publisher computes the global model $\bar{a_j}=\frac{Dec_{sk}(\Sigma_{i=1}^n c_{i,j})}{n}$ for $j=1, \ldots, l$ via the smart contract $SC$. 
}
	\item \textbf{Global Model Verification.}\\
	The publisher runs Algorithm \ref{verify sum}($S^3_1, \ldots, S^3_n,g_{pub},$ $ C_{i,1}, \ldots, C_{i,l}, g_{i,1}, \ldots, g_{i,l},\phi'_1, \ldots, \phi'_n, \bar{a_1}, \ldots,\bar{a_l}$) to check the correctness of the submitted models involved in the summation of the global model $\bar{a_1}, \ldots, \bar{a_l}$. 
	
\end{enumerate}

The completeness of the aggregation is guaranteed by equation \ref{aggregation} and the soundness relies on the $\Sigma$-protocol.
\begin{align}
    \label{aggregation}
    \bar{a_j}
    & = \frac{Dec_{sk}(\Sigma_{i=1}^n c_{i,j})}{n}\notag\\
    & = \frac{\Sigma_{i=1}^n a_{i,j}}{n}
      + \frac{\Sigma_{i=1}^n (s_{i,j}-s_{i-1,j})}{n}\notag\\
    & = \frac{\Sigma_{i=1}^n a_{i,j}}{n}
\end{align}

Specifically, the $\Sigma$-protocol here is to check if the equation \ref{sumsigma1} and equation \ref{sumsigma2} hold, where $e=Hash(g_{pub}||g_{i,j}||C_1||C_2||g_{pub}^{r_1}||g_{i,j}^{r_2})$.
\begin{align}
    \label{sumsigma1}
    g_{pub}^{r_1+ec_j} \equiv g_{pub}^{r_1}C_1^{e} \mod p\\
    \label{sumsigma2}
    g_{i,j}^{r_2+e(c_j+d_j)} \equiv g_{i,j}^{r_2}C_2^{e} \mod p
\end{align}
\textcolor{black}{As the global model summed, that the local model used is indeed the one submitted before can be checked with a proof generated by the trainer using Algorithm \ref{prove sum} and verified by the publisher using Algorithm \ref{verify sum}. 
And the correctness of the summation can be guaranteed by the smart contract. 
}
\begin{algorithm}
	\caption{{\rm \textbf{Gen\_Sum\_Prf(.)}}: Proof generation of the summation\label{prove sum}}
	\hspace*{0.07in}{\bf Input:}
	Statement $\phi_i=\{a_{i,j}\}_{j=1}^l$, t-list $T_i=\{t_{i,1}, \ldots, t_{i,l}\}$, generator $g_{pub}$.\\
	\hspace*{0.07in}{\bf Output:}
	The sigma proof $S^3_i$.
	\begin{algorithmic}[1]
		\For{$j$ in $1 \ldots l$}
		\State Set $c_j=a_{i,j}$, $d_j=t_{i,j}$, choose generator $g_{i,j}\in G$, $C_{i,j}=g_{pub}^{c_j}, C_2=g_{i,j}^{c_j+d_j}$
		\State Run $\Sigma$-protocol to obtain proof $s^3_{i,j}=PoK\{(c_j,d_j,$
		$r_1,r_2):C_{i,j} = g_{pub}^{c_j}, C_2 = g_{i,j}^{c_j+d_j}\}$
		\EndFor
		\\
		\Return $S^3_i=\{s^3_{i,1}, \ldots, s^3_{i,l}\}, g_{i,1}, \ldots, g_{i,l}, C_{i,1}, \ldots, C_{i,l}$
	\end{algorithmic}
\end{algorithm}

\begin{algorithm}
	\caption{{\rm \textbf{Vrf\_Sum\_Prf(.)}}: Proof verification of the summation\label{verify sum}}
	\hspace*{0.07in}{\bf Input:}
	Proof $S^3_1, \ldots, S^3_n$, generator $g_{pub}$, commitments $C_{1,1}, \ldots, C_{n,l}$, generators $g_{1,1}, \ldots, g_{n,l}$, the modified statement $\phi'_1, \ldots, \phi'_n$, the global result $\bar{a_1}, \ldots, \bar{a_l}$.\\
	\hspace*{0.07in}{\bf Output:}
	The verification result $b$.
	\begin{algorithmic}[1]
		\For{$j$ in $1 \ldots l$}
		\If{$\Pi_{i=1}^{n} C_{i,j}\neq g_{pub}^{n \bar{a_j}}$}
		\State \textbf{return} false
		\EndIf
		\For{$i$ in $1 \ldots n$}
		\State Compute $C_2=g_{i,j}^{a'_{i,j}}$
		\If{the verification of $s^3_{i,j}$ fails}
		\State \textbf{return} false
		\EndIf
		\EndFor
		\EndFor
		\\
		\Return true
	\end{algorithmic}
\end{algorithm}

\begin{figure*}
	\centering
	\includegraphics[width=6.5in]{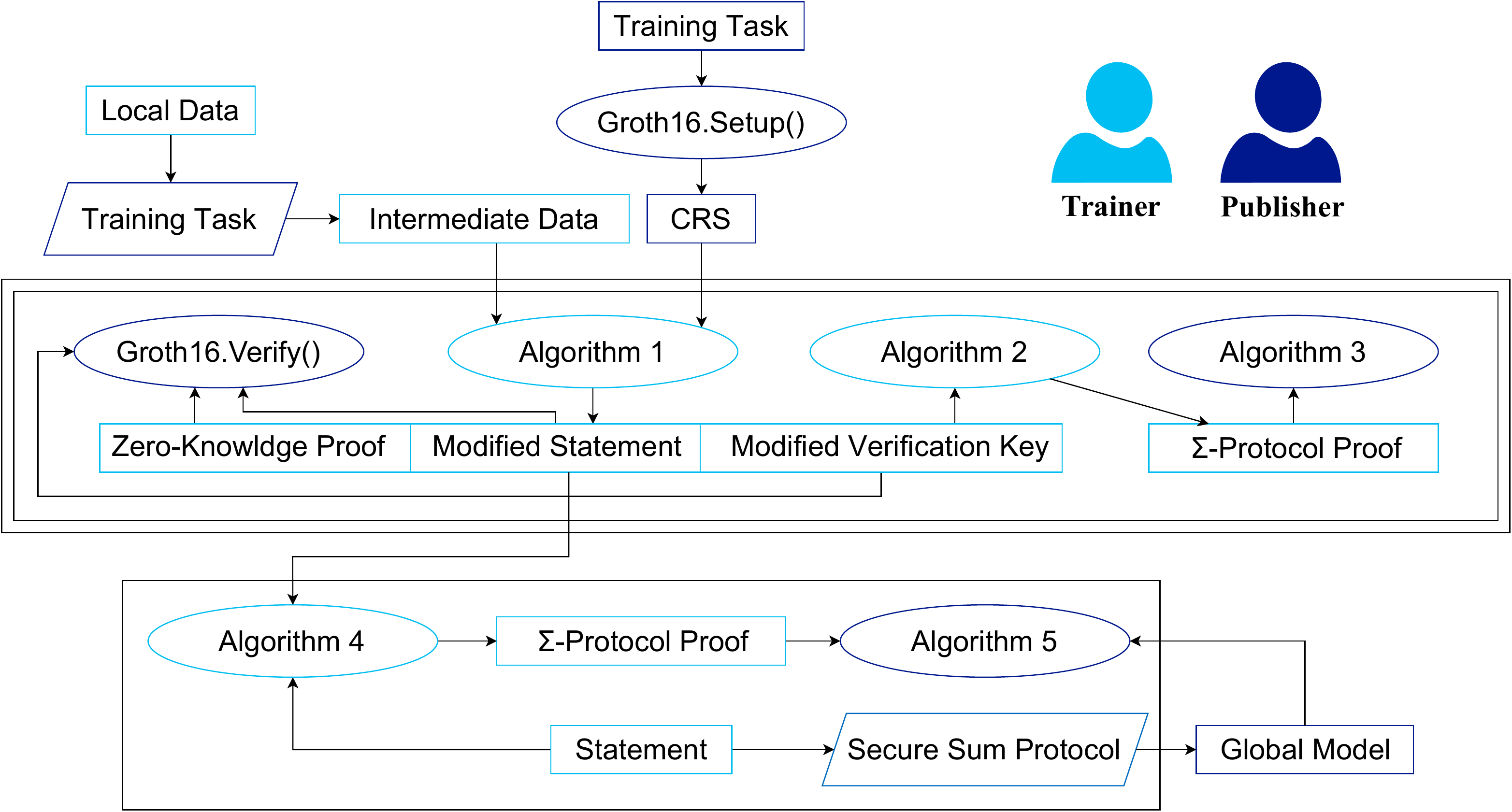}
	\caption{The architecture of the ZKP-FL scheme.}
	\label{algo}
\end{figure*}

Algorithms \ref{modify}, \ref{prove modification}, and \ref{verify modification} and $Groth16.Verify()$ ensure that the training process is carried out correctly. 
The secure sum protocol computes the global model without revealing the local one. 
Algorithms \ref{prove sum} and \ref{verify sum} ensure that the submitted models involved in the secure sum process is as claimed. 
The whole procedure is shown in Fig. \ref{algo}. 
As the training may consist of many rounds, the trainer has to generate ZKP for each round, which is circled by the double box. 
Nevertheless, these procedures can be carried out in parallel. 

%% file: practical.tex
There are still two practical problems in the ZKP-FL scheme, the fraction and integer mapping, and the non-linear operation. 
In this section, we further propose the Practical ZKP-FL scheme to solve these problems.

\subsection{The Fraction and Integer Mapping}
While handling the parameters in the ZKP-FL scheme, we observed that the majority of the parameters involved in ML algorithms were fractions. 
However, for ZKP, the parameters are merely integers. 
Thus, all of the fractions have to be mapped to the integer domain for supporting the execution of cryptographic operations based on elliptic curves. 
We use the scaling method described in Algorithm \ref{scale} to solve this problem. 
Precisely, trainers first reserve a certain decimal place for all fractions in the training process. 
The concrete decimal places are decided by the accuracy of the local training. 
Each trainer can reserve the appropriate decimal places according to the training process for their local data. 
Then, trainers can change the truncated fractions into integers by multiplying with a certain ratio $rat$. 
The ratio is chosen to be as large as possible to minimize the impact of truncation on accuracy. 

Due to the limitation of the number of bits, only numbers between $-2^{63}$ and $2^{63}-1$ can be used in the circuit. 
Considering the federated learning algorithm can be flattened into several binary expressions, such as $sym_i\ \bm{op}\ sym_j=sym_k$. 
Then, all these intermediate values $sym_1, sym_2, \ldots, sym_m$ can be obtained by training one time.
To ensure that no overflow occurs, we calculate $rat=arg\ max\left\{sym_{max}\cdot 10^{rat}\leq 2^{63}-1\ and\ sym_{min}\cdot 10^{rat}\geq \notag\right. \\ \left. -2^{63} :rat\in Z^*\right\}$. 
Then all the values can be expanded $10^{rat}$ times as $sym_1\cdot10^{rat},sym_2\cdot10^{rat}, \ldots, sym_m\cdot10^{rat}$, while all binary expressions still hold. 
With those integer intermediate values and binary expressions, the corresponding Quadratic Arithmetic Program (QAP) can be obtained and eventually generate arithmetic circuits for the generation of ZKPs.

\begin{algorithm}[htb]
	\caption{The fraction and integer mapping of the PZKP-FL scheme\label{scale}}
	\hspace*{0.07in}{\bf Input:}
	The training algorithm $F$, the local data $d$, original parameters $p$, training times $N$;\\
	\hspace*{0.07in}{\bf Output:}
	Integer intermediate values $sym_1,sym_2, \ldots, sym_m$.
	\begin{algorithmic}[1]
		\State Obtain intermediate parameters $sym_1,sym_2, \ldots, sym_m\leftarrow F(d,p,N)$
		\State Compute ratio $rat=arg\ max\{sym_{max}\cdot 10^{rat}\leq 2^{63}-1\ and\ sym_{min}\cdot 10^{rat}\geq -2^{63}:rat\in Z^*\}$
		\For{$i$ in $1 \ldots m$}
		\State $sym_i*=10^{rat}$
		\EndFor
		\\
		\Return $sym_1,sym_2, \ldots, sym_m$
	\end{algorithmic}
\end{algorithm}

\subsection{Non-linear Operations}
Non-linear operations are common in the federated learning algorithm. 
Here, we take the activation function of neural network as an example to demonstrate the proposed scheme can support non-linear operations. 
Most of the activation functions have relatively non-linear operations, such as \textit{Sigmoid}, and \textit{tanh} functions, which contain \textit{exp(x)}, \textit{sqrt(x)} operations. 
Such complex operations are not friendly to the languages describing arithmetic circuits. 
We approximate complex operations by linear operations through a Taylor expansion method described in Algorithm \ref{expansion} in the PZKP-FL scheme. 
The degree of the expansion is determined by a certain error, which reduces the impact of the approximation on the accuracy with shorter expression. 
We use the $i^{th}$ order Taylor expansion of the complex operation to as the simple operation, then evaluate the error between the simple operation and the original complex operation at $x$, until the error is less than a given threshold. 
The concrete value of the threshold is related with the specific federated learning tasks.

\begin{algorithm}[htb]
	\caption{Non-linear operation of the PZKP-FL scheme \label{expansion}}
	\hspace*{0.07in}{\bf Input:}
	The training algorithm $F$, the private input $x$, the error $E$, and the max value $MAX$.\\
	\hspace*{0.07in}{\bf Output:}
	The simple training algorithm $F$;
	\begin{algorithmic}[1]
		\For{complex operations $com$ in $F$}
		\State Get the Taylor expansion form $tay$ of the complex operation $com$;
		\State Get an empty simple expression $sim$
		\State $e=MAX$, $i=0$; 
		\While{$e>E$}
		\State $sim$ = $tay.i$ \# Take the taylor expansions of order $i$ as simple expressions 
		
		\State $e=|sim(x)-com(x)|$
		\EndWhile
		\State Replace the complex operation $com$ with simple operation $sim$
		\EndFor
		\\
		\Return $F$
	\end{algorithmic}
\end{algorithm}

In practice, in order to reduce the number of different circuits, we replace the same complex operation with the same Taylor expansion that meets the accuracy requirements for all values involved in the training, namely the Taylor expansion of the largest order that occurs. 
In this way, there would be only one circuit to be generated.

%% file: security.tex
\textcolor{black}{The PZKP-FL scheme has to attain three goals, the public verification of local computation, the public verification of global aggregation, and the privacy protection of local data, simultaneously.
In this section, we show the formal proofs for these goals.}

\subsection{The Proof for the Public Verification of Local Computation}

\begin{theorem}
In the PZKP-FL scheme, all of the trainers and the publisher can verify that all of the local parameters are computed based on the given AI model with local data.
Then for all PPT adversaries $A$, there is a negligible function $\mathrm{negl}$ such that: 

$\mathrm{Pr}[(\sigma, \tau)\leftarrow\mathrm{Setup}(R);(\phi, \pi, vk', s^1, s^2)\leftarrow A(R, \sigma) :\\ \phi \notin L_R, vk' \notin L_R \mathrm{\ and\ } Vfy(R, \sigma, vk', \phi, \pi, s^1, s^2) = 1 ] \leq \mathrm{negl}$

\begin{proof}(skeleton)
Consider the verification $Vfy$ according to algorithm \ref{verify modification} in two parts $Vfy1$ and $Vfy2$, aiming at verifying algorithm \ref{modify} and \ref{prove modification} separately.

Firstly, in the PZKP-FL scheme, the proof $\pi_i$ of Piece $i$ is generated by Algorithm \ref{modify}. 
\textcolor{black}{It argues for the statement that given the statement $\phi$ and the verification key $vk$, the AI model has to be executed correctly. 
This statement can be guaranteed, because the PZKP-FL scheme uses the Algorithm $Groth16.Verify()$.
In other word, the ZKP of this statement verifies the satisfiability of an arithmetic circuit.
That is, given the inputs, outputs, and constraints, the computation of the arithmetic circuit has to be correct, suppose the output of the algorithm $Groth16.Verify()$ is true.}
Therefore, since the arithmetic circuit is converted from the training algorithm, and $\pi_i$ is verified, the satisfiability of the arithmetic circuit guarantees the correct execution of the AI model.
Which is to say, 

$\mathrm{Pr}[(\sigma, \tau)\leftarrow\mathrm{Setup}(R);(\phi, \pi, vk')\leftarrow A(R, \sigma) :
\phi \notin L_R \mathrm{\ and\ } Vfy1(R, vk', \phi, \pi) = 1] \leq \mathrm{negl}$

Secondly, the $\Sigma$ proof $s^1_i$ for the Piece $i$ is generated by the Algorithm \ref{prove modification}, proving the statement that 
given the verification key $vk$ and the modified verification key $vk'$, $vk'$ is a valid modification of $vk$ based on the Algorithm \ref{modify}, and the trainer has the knowledge of the modification at the same time. 
This is due to the fact that given the commitment $C=\Pi_{i=1}^{n}g_i^{c_i}$ and generators $g_1, \ldots, g_n$, a $\Sigma$-protocol can prove the knowledge of $c_1, \ldots, c_n$.

Thirdly, the $\Sigma$ proof $s^2_i$ for Pieces $i$ and $i-1$ is generated by the Algorithm \ref{prove modification}, proving the statement that the Piece $i$'s inputs equal to the Piece $i-1$'s outputs''. 
This is due to the fact that given commitments $C_1=\Pi_{i=1}^{n}g_{1,i}^{c_i}, C_2=\Pi_{i=1}^{n}g_{2,i}^{c_i}$ and generators $g_{1,1}, \ldots, g_{1,n},g_{2,1}, \ldots, g_{2,n}$, a $\Sigma$-protocol can prove that 
the knowledge of $c_1, \ldots, c_n$ and $C_1$ and $C_2$ commit to the same value $c_i$.
Which is to say,

$\mathrm{Pr}[(\sigma, \tau)\leftarrow\mathrm{Setup}(R);(vk', s^1, s^2)\leftarrow A(R, \sigma) :vk' \notin L_R \mathrm{\ and\ } Vfy2(R, \sigma, vk', s^1, s^2) = 1 ] \leq \mathrm{negl}$

\textcolor{black}{When combining the above three conditions and run this proof for all rounds, the statement has to be correct in the whole training process.
We finish the proof.}
$\hfill\blacksquare$
\end{proof}
\end{theorem}

\subsection{The Proof for the Public Verification of Global Aggregation}
\begin{theorem}
In the PZKP-FL scheme, all of the trainers and the publisher can verify that the aggregated parameters are computed according to the sum protocol with local parameters.

\begin{proof}(skeleton)
\textcolor{black}{Firstly, the specific process of parameter aggregation is executed on a smart contract. 
The computation is carried out by miners according to the given smart contract, and the correctness can be checked by any user available to the blockchain. 
So it is negligible that the computation result can be changed, due to the soundness of the smart contract and the consensus mechanism used in the blockchain system. 
}

\textcolor{black}{Secondly, the $\Sigma$ proof $S^3_i$ for Piece $i$ is generated by Algorithm \ref{prove sum}, proving the statement that 
the commitment $C_i$ based on given generator $g_{pub}$ commits to the value $\phi$ used in the modified statement $\phi'_i$. }
This is due to the fact that given commitments $C_1=g_1^c, C_2=g_2^{c+d}$ and generators $g_1,g_2$, a $\Sigma$-protocol can prove the knowledge of $c,d$, and $C_1$ is a commitment to $c$; meanwhile, $C_2$ is a commitment to $c+d$. 
The above statement connects the value $\phi_i'$ submitted as the training result and the value $g_{pub}^{\phi_i}$ submitted in a secure sum protocol, indicating these two values contain the same result $\phi_i$.
The publisher can verify the aggregated parameter $\bar{\phi}$ by checking the equation $\Pi_{i=1}^{n} g_{pub}^{\phi_i} = g_{pub}^{n \bar{a_j}}$. 
If the equation holds, then the aggregated parameter must be computed from the submitted modified statement $\phi'_i$.

\textcolor{black}{When combining the above two conditions and run this proof for all rounds, the statement has to be correct in the whole training process.
We finish the proof.}
$\hfill\blacksquare$
\end{proof}
\end{theorem}

\subsection{The Proof for the Privacy Protection of Local Data}
\begin{theorem}
\textcolor{black}{In the PZKP-FL scheme, all of the participants are negligible to obtain any other trainer's local data and the plaintext of the intermediate parameters generated during in training process.}

\begin{proof}(skeleton)
\textcolor{black}{Firstly, in the PZKP-FL scheme, the data transmitted by a trainer $i$ is the modified statement $\phi'_i$ and the encrypted statement $c_i$. 
$\phi'_i$ is noised by a uniformly distributed random number $t_i$ as $\phi'_i=\phi_i+t_i$. 
Therefore, as long as the random number $t_i$ cannot be correctly guessed, $\phi'_i$ is secure. 
Here, the probability to correctly guess $t_i$ is definitely negligible.}

\textcolor{black}{Secondly, since $c_i$ is encrypted with the public key $pk$ generated by the publisher, and it can only be decrypted by the publisher.
Thus, as long as the encryption scheme is secure, $c_i$ is secure to other trainers. }

\textcolor{black}{Thirdly, $c_i$ is also an encryption of $\phi_i+a-b$, and both $a, b$ are uniformly distributed random numbers and known only by two trainers.
Therefore, as long as the random number $a, b$ cannot be correctly guessed, $c_i$ is secure to the publisher.
Here,  the probability to correctly guess $a, b$ is also negligible.}

\textcolor{black}{Finally, all of the proofs are zero knowledge because they are generated by using zk-SNARKs and $\Sigma$-protocols.
The concrete steps are listed as below.}

\begin{strip}
\begin{tcolorbox}[colback=white,colframe=black]
\textbf{Common inputs}: The proving key $pk$ and the verification key $vk$.

\textbf{Trainer $i$'s input}: Local data $d$.

\quad The trainer runs $Groth16.Prove()$ and Algorithm \ref{modify} to obtain ZKP $\pi_i$, the modified verification key $vk'_i$ and the modified statement $\phi'_i$. 
The trainer runs Algorithm \ref{prove modification} to obtain the $\Sigma$ proof $s^1_i$ and $s^2_i$. 
The trainer runs Algorithm \ref{prove sum} to obtain the $\Sigma$ proof $s^3_i$.

1. Show that $\pi_j$ is a valid ZKP under the verification key $vk'_j$ and the statement $\phi'_j$.

$$PoK\{(\pi_j,vk'_j,\phi'_j): Groth16.Verify(\pi_j,vk'_j,\phi'_j)=1\}$$

2. Show that $\phi'_j$ is a valid modification on $\phi_j$ and $vk'_j$ is a valid modification on $vk_j$.

$$PoK\{(vk_j,vk'_j,s^1_j):\Pi_{k=1}^l vk_j.\gamma\_abc_k^{s^1_j.z_k}=s^1_j.a (vk_j'.\gamma\_abc_0 \cdot vk_j.\gamma\_abc_0^{-1})^{s^1_j.e}\mod p\}$$

3. Show that $\phi'_j$ and $\phi'_{j-1}$ are modified in the same way.

$$PoK\{(vk_j, vk'_j, tsum^1_j, tsum^2_{j-1}, s^2_j):\Pi_{j=1}^{l/2}vk_j.\gamma\_abc_k^{s^2_j.z_j}=s^2_j.a_1(tsum^1_j)^{s^2_j.e}\mod p \ \wedge$$
$$ \Pi_{j=l/2+1}^{l}vk_j.\gamma\_abc_k^{s^2_j.z_k}=s^2_j.a_2(tsum^2_{j-1})^{s^2_j.e}\mod p\}$$

4. Show that the global parameters $\bar{x}$ is exactly the sum of the value $x_i$ that trainers submitted as $x_i+t_i$ before.

$$PoK\{(g_{pub}, g_{pub}^x, \bar{x}, g_i, C_i, s^3_j, x_i+t_i):\Pi_{i=1}^{n} g_{pub}^{x_i}=g_{pub}^{n\bar{x}} \wedge$$
$$ g_{pub}^{s^3_i.z_1} = s^3_i.a_1(s^3_i.C_i)^{s^3_i.e} \wedge g_i^{s^3_i.z_1+s^3_i.z_2} = s^3_i.a_2\cdot s^3_i.a_3 (g_i^{x_i+t_i})^{s^3_i.e}\}$$

\end{tcolorbox}
\end{strip}

\textcolor{black}{All in all, when combing the above four conditions and run this proof for all rounds, the statement has to be correct in the whole training process.}
$\hfill\blacksquare$
\end{proof}
\end{theorem}

%% file: performance.tex
\begin{table*}
	\centering
	\caption{The experimental results of a classification task and a prediction task.}
	\label{exp1t}
	\begin{tabular}{c|ccccc}
	    \hline
	    \textbf{\thead{Task}} & \textbf{\thead{Setup Time\\(s)}} & \textbf{\thead{Proof Generation Time\\(per proof) (s)}} & \textbf{\thead{Proof Verification Time\\(per proof) (s)}} & \textbf{\thead{Number of Proof}} & \textbf{\thead{Accuracy}}
		\\ \hline
		\thead{Iris classification} & \thead{950.34} & \thead{89.99} & \thead{25.75} & \thead{90,000} & \thead{96\%}
		\\ \hline
		\thead{Price prediction} & \thead{514.19} & \thead{50.09} & \thead{1.55} & \thead{500} & \thead{88\%}
	\end{tabular}
\end{table*}

\begin{table*}
	\centering
	\caption{The experimental results of the impact of the piece size.}
	\label{exp3t}
		\begin{tabular}{c|cccccc}
			\textbf{\thead{Piece Size (number of epoch)}} & 1 & 2 & 3 & 5 & 10 & 15
			\\ \hline
			\textbf{\thead{Setup Time (s)}} & \thead{508.25} & \thead{966.43} & \thead{1518.59} & \thead{2622.10} & \thead{5247.52} & \thead{7268.61}
			\\ \hline
			\textbf{\thead{Proof generation (per proof) (s)}} & \thead{50.74} & \thead{95.95} & \thead{140.70} & \thead{225.10} & \thead{439.16} & \thead{533.24}
			\\ \hline
			\textbf{\thead{Proof verification (per proof) (s)}} & \thead{1.56} & \thead{1.54} & \thead{1.19} & \thead{1.15} & \thead{1.13} & \thead{1.12}
			\\ \hline \hline
			\textbf{\thead{Circuit Constraints}} & \thead{973,617} & \thead{1,923,246} & \thead{2,873,142} & \thead{4,773,477} & \thead{9,525,930} & \thead{14,279,460}
			\\ \hline
			\textbf{\thead{Points in CRS}} & \thead{968,155} & \thead{1,912,324} & \thead{2,856,760} & \thead{4,746,175} & \thead{9,471,328} & \thead{14,197,558}
			\\ \hline \hline
			\textbf{\thead{Arithmetic Circuit Size (KB)}} & \thead{162,612} & \thead{323,291} & \thead{484,035} & \thead{805,678} & \thead{1,610,476} &\thead{2,416,076}
			\\ \hline
			\textbf{\thead{Proving Key Size (KB)}} & \thead{347,279} & \thead{687,122} & \thead{1,092,583} & \thead{1,903,677} & \thead{3,800,842} & \thead{5,674,057}
			\\ \hline
			\textbf{\thead{Verification Key Size (per proof) (KB)}} & \thead{2} & \thead{2} & \thead{2} & \thead{2} & \thead{2} & \thead{2}
			\\ \hline
			\textbf{\thead{Proof Size (per proof) (KB)}} & \thead{5} & \thead{5} & \thead{5} & \thead{5} & \thead{5} & \thead{5}
			\\ \hline \hline
		\end{tabular}
\end{table*}

To verify the feasibility and performance of our scheme, we selected two specific ML tasks, namely the iris classification task and the house price prediction task. 
There are two kinds of participants in this experiment, the publisher and the trainer. 
All the evaluations run on an Ubuntu 18.04 instance with 4GiB RAM and Intel Xeon Gold 5128 CPU.
The ZKP toolbox we used is ZoKrates, which can convert codes into arithmetic circuit constraints and generate ZKPs based on inputs.

The experiment is carried out as follows. 
First, the publisher translates the original ML algorithm into a integer-version one using the method described in Section \ref{PZKP} and splits it into pieces. 
Then, the publisher runs the $Groth16.Setup()$ algorithm to generate the corresponding proving key and verification key. All of these are sent to each trainer.
As the trainers carry out the training process with the integer algorithm and preserve all the intermediate data, they can generate the ZKP by using the algorithms described in Section \ref{ZKP}. 
Once the training is completed, the trainer sends all the proofs with the local model to the publisher. 
As proofs are verified, the publisher and trainers run the secure sum protocol to aggregate the global model. 
Finally, the global model is verified. 
The experimental parameters involved in the above procedure are shown in Table \ref{param}. 
We take the iris classification task as an example to illustrate the experiment process. 
In the iris classification task, we have 75 training samples, the model is trained 1,200 times. 
Regarding each sample in each iteration as one piece, then we have 90,000 pieces in total. 
The whole training procedure is expanded by $10^7$ as the
scaling ratio. 
We choose Sigmoid as the activation function, \ie $f(x)=\frac{1}{1+e^{-x}}$. 
The $i^{th}$ order Taylor expansion of sigmoid function takes the form as $$f(x)\approx \frac{1}{1+(1+(-x)+\frac{(-x)^2}{2!}+\ldots+\frac{(-x)^i}{i!})}$$. 
Considering the error between the approximate function and the original function as 0.0001, we set the approximate function to the fifth-order Taylor expansion. 
Next, we convert a training process in fraction into integer form. 
We run the integer version training algorithm, while saving all the intermediate values and the local result. 
The accuracy of the local result is 96\%, which is the same as the original fraction version.
As each piece contains the same computational operation, we compile only one piece of code into ZoK language. 
With the arithmetic circuit generated by ZoKrates through the ZoK code, the setup in ZKP is complete.
Then, we can generate the proof of correctness of the computation process in each piece through Groth16 and Algorithm \ref{modify}, with inputs and outputs of each piece saved while training. 
Using Algorithm \ref{prove modification}, we can generate the ZKP based on $\Sigma$-protocol that connects inputs and outputs between pieces. 
Finally, we generate 90,000 ZKPs about the correctness of the whole training process. 
It is worth noting that the splitting method is determined based on the memory required for the initialization operation. 
The splitting method given in the following table allows the publisher to complete the setup within a memory size of 4GB.

\begin{table}[htbp]
	\centering
	\caption{The experimental parameters.}
	\label{param}
	\begin{tabular}{c|cc}
            \hline
            \textbf{\thead{Task}} & \thead{Iris classification} & \thead{Price prediction} \\ \hline
            \textbf{\thead{Splitting Method}} & \thead{Each sample\\in each round} & \thead{Each round} \\
            \textbf{\thead{Expansion Ratio}} & \thead{10,000,000} & \thead{10,000,000} \\
            \textbf{\thead{Training Sample}} & \thead{75} & \thead{90} \\
            \textbf{\thead{Test Sample}} & \thead{75} & \thead{10} \\
            \textbf{\thead{Training Rounds}} & \thead{1,200} & \thead{500}
            
	\end{tabular}
\end{table}

The experimental results are shown in Table \ref{exp1t}. 
Given the secure sum protocol is an interactive process and consumes less time and computation, the experiment only counts the time spent on the setup, proof generation, and proof verification parts. 
If we train the model for the iris classification task without splitting the training algorithm, then during the proof generation process, the memory required to perform the setup will be 30 TiB at least, while the runtime is nearly 1000 days, which is completely unrealistic. 
Even if the required arithmetic circuit constraints are obtained, it takes more than 90 days to generate the proofs and more than 25 days to verify them, which is not practical. 
Splitting not only makes setup operations practical, but also allows both proof generation and verification operations to be executed in parallel. 
The time required for all proof generation and verification operations can be reduced by parallel execution on multiple processors.

\begin{figure}
	\centering
	\includegraphics[width=0.4\textwidth]{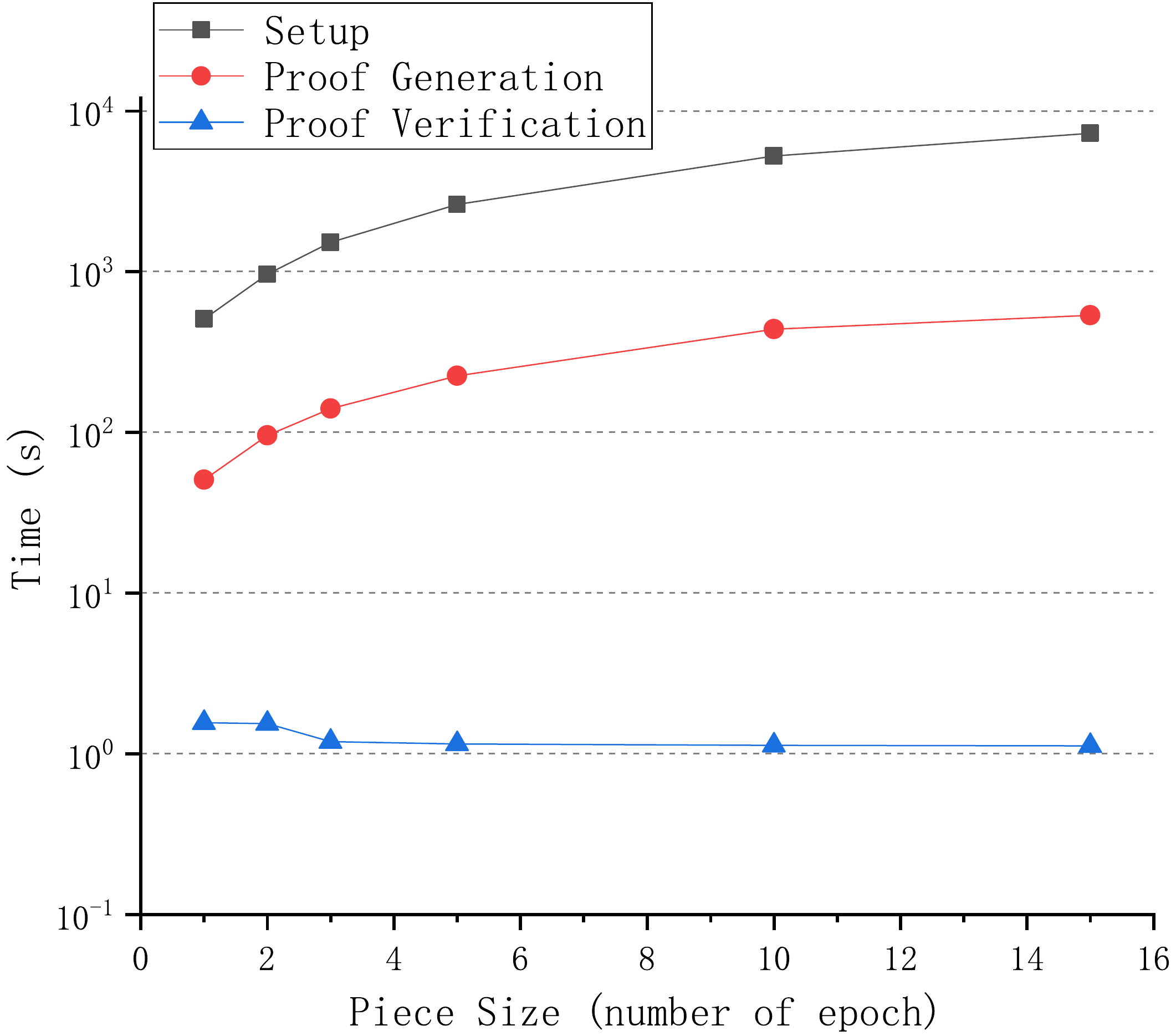}
	\caption{Impact of piece size on the running time of ZKP.} \label{exp3f1}
\end{figure}

\begin{figure}
	\centering
	\includegraphics[width=0.4\textwidth]{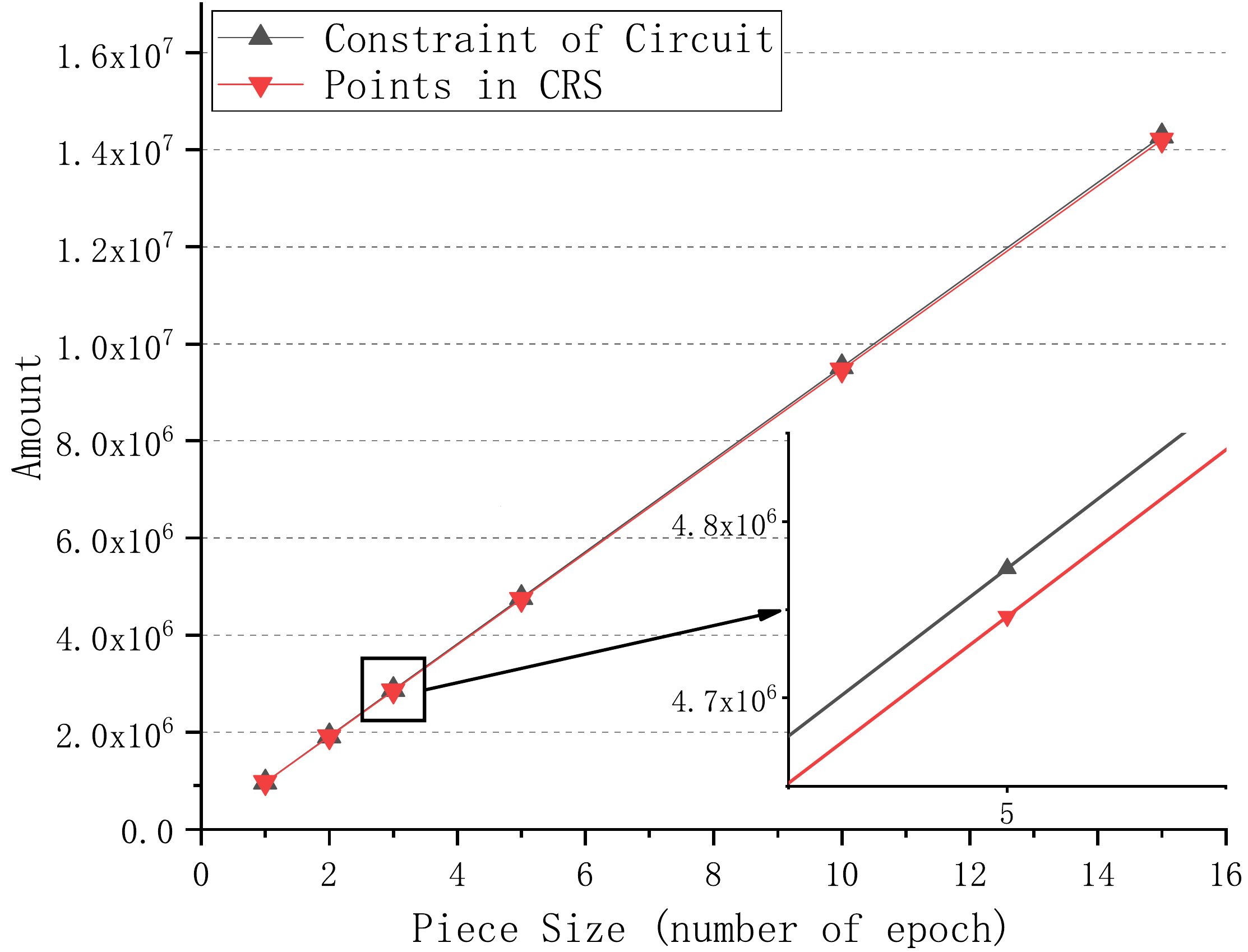}
	\caption{Impact of piece size on the ZKP circuit.} \label{exp3f2}
\end{figure}

\begin{figure}
	\centering
	\includegraphics[width=0.4\textwidth]{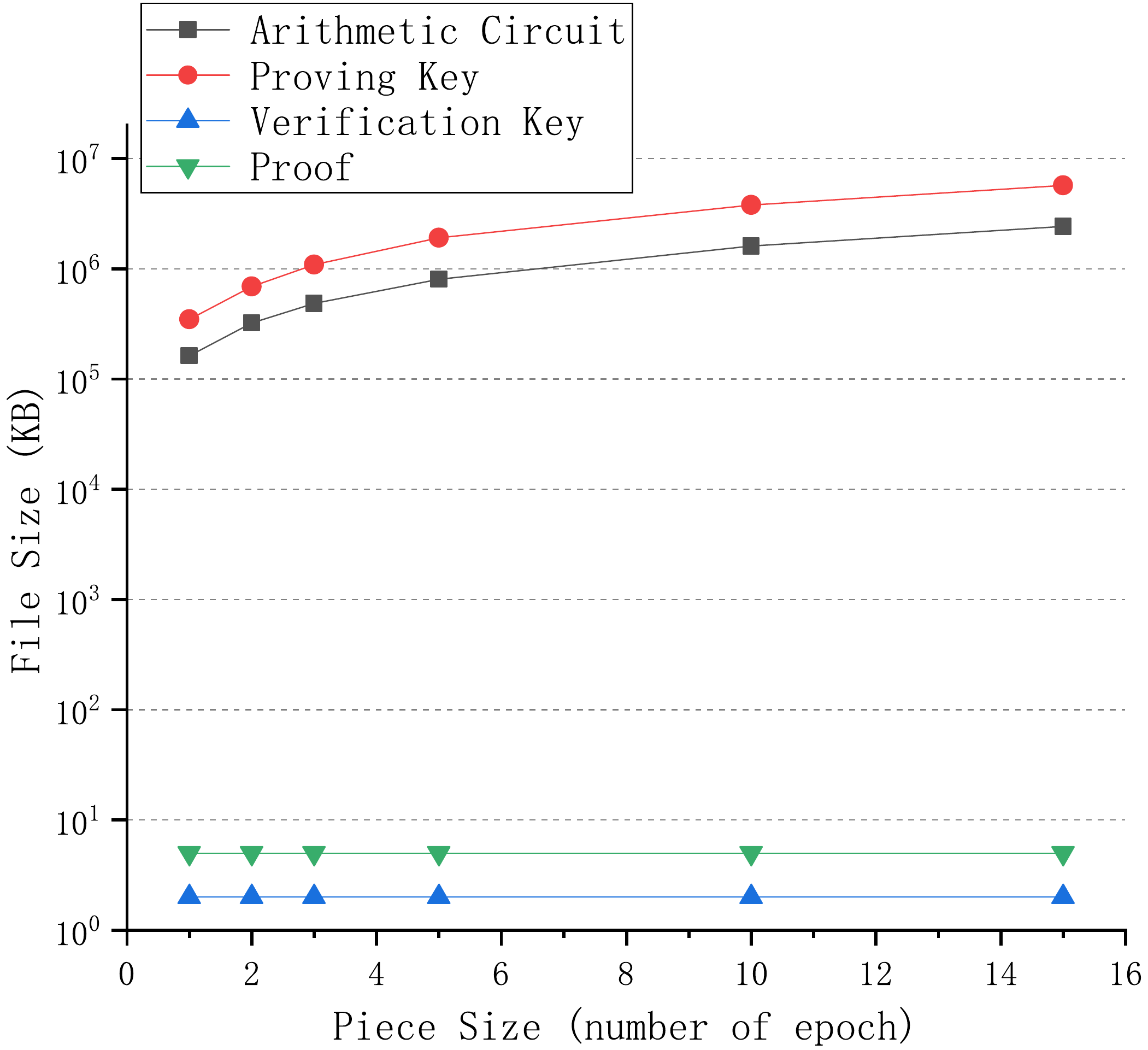}
	\caption{Impact of piece size on file size generated in the ZKP process.} \label{exp3f3}
\end{figure}

Another experiment explores the relationship between the size of one piece and the running cost. 
Take the price prediction task as an example. 
The entire model needs to be trained for 500 rounds. 
We define the piece size that splits each round as a piece as 1. 
We take 5 different piece sizes, including 1, 2, 3, 5, and 10.
The running cost is shown in Table \ref{exp3t} and Figs. \ref{exp3f1}, \ref{exp3f2}, and \ref{exp3f3}. 
The circuit size (\ie a set of constraints) is proportional to the size of piece. 
It influences the circuit generation time (setup), proof generation time (per proof), file size of proving key and circuit, which are all proportional to the size of piece. 
The verification time (per proof), the size of file for verification (verification key, proof) is constant and relatively smaller. 
The overall proof generation time is effected by both the size and the number of pieces.
However, as the proof generation can run in parallel, the smaller the piece size, the shorter the overall running time will be for trainers.

%% file: main.bbl
\begin{thebibliography}{10}
\providecommand{\url}[1]{#1}
\csname url@samestyle\endcsname
\providecommand{\newblock}{\relax}
\providecommand{\bibinfo}[2]{#2}
\providecommand{\BIBentrySTDinterwordspacing}{\spaceskip=0pt\relax}
\providecommand{\BIBentryALTinterwordstretchfactor}{4}
\providecommand{\BIBentryALTinterwordspacing}{\spaceskip=\fontdimen2\font plus
\BIBentryALTinterwordstretchfactor\fontdimen3\font minus
  \fontdimen4\font\relax}
\providecommand{\BIBforeignlanguage}[2]{{%
\expandafter\ifx\csname l@#1\endcsname\relax
\typeout{** WARNING: IEEEtran.bst: No hyphenation pattern has been}%
\typeout{** loaded for the language `#1'. Using the pattern for}%
\typeout{** the default language instead.}%
\else
\language=\csname l@#1\endcsname
\fi
#2}}
\providecommand{\BIBdecl}{\relax}
\BIBdecl

\bibitem{esteva2017dermatologist}
A.~Esteva, B.~Kuprel, R.~A. Novoa, J.~Ko, S.~M. Swetter, H.~M. Blau, and
  S.~Thrun, ``Dermatologist-level classification of skin cancer with deep
  neural networks,'' \emph{nature}, vol. 542, no. 7639, pp. 115--118, 2017.

\bibitem{lampos2021artificial}
V.~Lampos, J.~Mintz, and X.~Qu, ``An artificial intelligence approach for
  selecting effective teacher communication strategies in autism education,''
  \emph{npj Science of Learning}, vol.~6, no.~1, pp. 1--10, 2021.

\bibitem{zhu2019deep}
L.~Zhu, Z.~Liu, and S.~Han, ``Deep leakage from gradients,'' \emph{Advances in
  Neural Information Processing Systems}, vol.~32, 2019.

\bibitem{lam2021gradient}
M.~Lam, G.-Y. Wei, D.~Brooks, V.~J. Reddi, and M.~Mitzenmacher, ``Gradient
  disaggregation: Breaking privacy in federated learning by reconstructing the
  user participant matrix,'' in \emph{International Conference on Machine
  Learning}.\hskip 1em plus 0.5em minus 0.4em\relax PMLR, 2021, pp. 5959--5968.

\bibitem{hamer2020fedboost}
J.~Hamer, M.~Mohri, and A.~T. Suresh, ``Fedboost: A communication-efficient
  algorithm for federated learning,'' in \emph{International Conference on
  Machine Learning}.\hskip 1em plus 0.5em minus 0.4em\relax PMLR, 2020, pp.
  3973--3983.

\bibitem{liu2020fedvision}
Y.~Liu, A.~Huang, Y.~Luo, H.~Huang, Y.~Liu, Y.~Chen, L.~Feng, T.~Chen, H.~Yu,
  and Q.~Yang, ``{Fedvision}: An online visual object detection platform
  powered by federated learning,'' in \emph{Proceedings of the AAAI Conference
  on Artificial Intelligence}, vol.~34, no.~08, 2020, pp. 13\,172--13\,179.

\bibitem{blum2021one}
A.~Blum, N.~Haghtalab, R.~L. Phillips, and H.~Shao, ``One for one, or all for
  all: Equilibria and optimality of collaboration in federated learning,'' in
  \emph{International Conference on Machine Learning}.\hskip 1em plus 0.5em
  minus 0.4em\relax PMLR, 2021, pp. 1005--1014.

\bibitem{sun2020federated}
L.~Sun and L.~Lyu, ``Federated model distillation with noise-free differential
  privacy,'' \emph{arXiv preprint arXiv:2009.05537}, 2020.

\bibitem{sun2020ldp}
L.~Sun, J.~Qian, and X.~Chen, ``{LDP-FL}: Practical private aggregation in
  federated learning with local differential privacy,'' \emph{arXiv preprint
  arXiv:2007.15789}, 2020.

\bibitem{kairouz2021distributed}
P.~Kairouz, Z.~Liu, and T.~Steinke, ``The distributed discrete {Gaussian}
  mechanism for federated learning with secure aggregation,'' in
  \emph{International Conference on Machine Learning}.\hskip 1em plus 0.5em
  minus 0.4em\relax PMLR, 2021, pp. 5201--5212.

\bibitem{wu2022adaptive}
X.~Wu, Y.~Zhang, M.~Shi, P.~Li, R.~Li, and N.~N. Xiong, ``An adaptive federated
  learning scheme with differential privacy preserving,'' \emph{Future
  Generation Computer Systems}, vol. 127, pp. 362--372, 2022.

\bibitem{li2020privacy}
Y.~Li, Y.~Zhou, A.~Jolfaei, D.~Yu, G.~Xu, and X.~Zheng, ``Privacy-preserving
  federated learning framework based on chained secure multiparty computing,''
  \emph{IEEE Internet of Things Journal}, vol.~8, no.~8, pp. 6178--6186, 2020.

\bibitem{asad2021ceep}
M.~Asad, A.~Moustafa, and M.~Aslam, ``{CEEP-FL}: A comprehensive approach for
  communication efficiency and enhanced privacy in federated learning,''
  \emph{Applied Soft Computing}, vol. 104, p. 107235, 2021.

\bibitem{ruckel2022fairness}
T.~R{\"u}ckel, J.~Sedlmeir, and P.~Hofmann, ``Fairness, integrity, and privacy
  in a scalable blockchain-based federated learning system,'' \emph{Computer
  Networks}, vol. 202, p. 108621, 2022.

\bibitem{voigt2017eu}
P.~Voigt and A.~Von~dem Bussche, ``The {EU General Data Protection Regulation
  (GDPR)},'' \emph{A Practical Guide, 1st Ed., Cham: Springer International
  Publishing}, vol.~10, no. 3152676, pp. 10--5555, 2017.

\bibitem{de2018guide}
L.~de~la Torre, ``A guide to the {California} consumer privacy act of 2018,''
  \emph{Available at SSRN 3275571}, 2018.

\bibitem{determann2021china}
L.~Determann, Z.~J. Ruan, T.~Gao, and J.~Tam, ``China’s draft personal
  information protection law,'' \emph{Journal of Data Protection \& Privacy},
  vol.~4, no.~3, pp. 235--259, 2021.

\bibitem{groth2016size}
J.~Groth, ``On the size of pairing-based non-interactive arguments,'' in
  \emph{Annual international conference on the theory and applications of
  cryptographic techniques}.\hskip 1em plus 0.5em minus 0.4em\relax Springer,
  2016, pp. 305--326.

\bibitem{goldwasser1989knowledge}
S.~Goldwasser, S.~Micali, and C.~Rackoff, ``The knowledge complexity of
  interactive proof systems,'' \emph{SIAM Journal on computing}, vol.~18,
  no.~1, pp. 186--208, 1989.

\bibitem{groth2010short}
J.~Groth, ``Short non-interactive zero-knowledge proofs,'' in
  \emph{International Conference on the Theory and Application of Cryptology
  and Information Security}.\hskip 1em plus 0.5em minus 0.4em\relax Springer,
  2010, pp. 341--358.

\bibitem{damgaard2002sigma}
I.~Damg{\aa}rd, ``On $\sigma$-protocols,'' \emph{Lecture Notes, University of
  Aarhus, Department for Computer Science}, p.~84, 2002.

\bibitem{he2005face}
X.~He, S.~Yan, Y.~Hu, P.~Niyogi, and H.-J. Zhang, ``Face recognition using
  {Laplacianfaces},'' \emph{IEEE transactions on pattern analysis and machine
  intelligence}, vol.~27, no.~3, pp. 328--340, 2005.

\bibitem{shepitsen2008personalized}
A.~Shepitsen, J.~Gemmell, B.~Mobasher, and R.~Burke, ``Personalized
  recommendation in social tagging systems using hierarchical clustering,'' in
  \emph{Proceedings of the 2008 ACM conference on Recommender systems}, 2008,
  pp. 259--266.

\bibitem{shi2016edge}
W.~Shi, J.~Cao, Q.~Zhang, Y.~Li, and L.~Xu, ``Edge computing: Vision and
  challenges,'' \emph{IEEE internet of things journal}, vol.~3, no.~5, pp.
  637--646, 2016.

\bibitem{mcmahan2017communication}
B.~McMahan, E.~Moore, D.~Ramage, S.~Hampson, and B.~A. y~Arcas,
  ``Communication-efficient learning of deep networks from decentralized
  data,'' in \emph{Artificial intelligence and statistics}.\hskip 1em plus
  0.5em minus 0.4em\relax PMLR, 2017, pp. 1273--1282.

\bibitem{huang2019patient}
L.~Huang, A.~L. Shea, H.~Qian, A.~Masurkar, H.~Deng, and D.~Liu, ``Patient
  clustering improves efficiency of federated machine learning to predict
  mortality and hospital stay time using distributed electronic medical
  records,'' \emph{Journal of biomedical informatics}, vol.~99, p. 103291,
  2019.

\bibitem{yang2019federated}
Q.~Yang, Y.~Liu, T.~Chen, and Y.~Tong, ``Federated machine learning: Concept
  and applications,'' \emph{ACM Transactions on Intelligent Systems and
  Technology (TIST)}, vol.~10, no.~2, pp. 1--19, 2019.

\bibitem{bonawitz2017practical}
K.~Bonawitz, V.~Ivanov, B.~Kreuter, A.~Marcedone, H.~B. McMahan, S.~Patel,
  D.~Ramage, A.~Segal, and K.~Seth, ``Practical secure aggregation for
  privacy-preserving machine learning,'' in \emph{proceedings of the 2017 ACM
  SIGSAC Conference on Computer and Communications Security}, 2017, pp.
  1175--1191.

\bibitem{luo2021feature}
X.~Luo, Y.~Wu, X.~Xiao, and B.~C. Ooi, ``Feature inference attack on model
  predictions in vertical federated learning,'' in \emph{2021 IEEE 37th
  International Conference on Data Engineering (ICDE)}.\hskip 1em plus 0.5em
  minus 0.4em\relax IEEE, 2021, pp. 181--192.

\bibitem{CCPA}
``{California Consumer Privacy Act (CCPA)},'' https://oag.ca.gov/privacy/ccpa,
  2022/04/02.

\bibitem{PIPL}
``Personal information protection law of the {People's Republic of China},''
  http://www.xinhuanet.com/politics/2021-08/20/c\_1127781552.htm, 2022/04/02.

\bibitem{liu2020federated}
F.~Liu, X.~Wu, S.~Ge, W.~Fan, and Y.~Zou, ``Federated learning for
  vision-and-language grounding problems,'' in \emph{Proceedings of the AAAI
  Conference on Artificial Intelligence}, vol.~34, no.~07, 2020, pp.
  11\,572--11\,579.

\bibitem{weng2021mystique}
C.~Weng, K.~Yang, X.~Xie, J.~Katz, and X.~Wang, ``Mystique: Efficient
  conversions for $\{$Zero-Knowledge$\}$ proofs with applications to machine
  learning,'' in \emph{30th USENIX Security Symposium (USENIX Security 21)},
  2021, pp. 501--518.

\bibitem{liu2021zkcnn}
T.~Liu, X.~Xie, and Y.~Zhang, ``{ZkCNN}: Zero knowledge proofs for
  convolutional neural network predictions and accuracy,'' in \emph{Proceedings
  of the 2021 ACM SIGSAC Conference on Computer and Communications Security},
  2021, pp. 2968--2985.

\bibitem{zhang2020zero}
J.~Zhang, Z.~Fang, Y.~Zhang, and D.~Song, ``Zero knowledge proofs for decision
  tree predictions and accuracy,'' in \emph{Proceedings of the 2020 ACM SIGSAC
  Conference on Computer and Communications Security}, 2020, pp. 2039--2053.

\bibitem{abadi2016deep}
M.~Abadi, A.~Chu, I.~Goodfellow, H.~B. McMahan, I.~Mironov, K.~Talwar, and
  L.~Zhang, ``Deep learning with differential privacy,'' in \emph{Proceedings
  of the 2016 ACM SIGSAC conference on computer and communications security},
  2016, pp. 308--318.

\bibitem{facebook}
T.~A. Press, ``Judge approves \$650m {Facebook} privacy lawsuit settlement,''
  https://apnews.com/article/technology-business-san-francisco-chicago-lawsuits-af6b42212e43be1b63b5c290eb5bfd85,
  2022/05/31.

\bibitem{equifax}
F.~T. Commission, ``Equifax to pay \$575 million as part of settlement with
  {FTC, CFPB}, and states related to 2017 data breach,''
  https://www.ftc.gov/news-events/news/press-releases/2019/07/equifax-pay-575-million-part-settlement-ftc-cfpb-states-related-2017-data-breach,
  2022/05/31.

\bibitem{nakamoto2008bitcoin}
S.~Nakamoto, ``Bitcoin: A peer-to-peer electronic cash system,''
  \emph{Decentralized Business Review}, p. 21260, 2008.

\bibitem{szabo1997formalizing}
N.~Szabo, ``Formalizing and securing relationships on public networks,''
  \emph{First monday}, 1997.

\bibitem{choi2013multi}
S.~G. Choi, J.~Katz, R.~Kumaresan, and C.~Cid, ``Multi-client non-interactive
  verifiable computation,'' in \emph{Theory of Cryptography Conference}.\hskip
  1em plus 0.5em minus 0.4em\relax Springer, 2013, pp. 499--518.

\bibitem{zhao2021veriml}
L.~Zhao, Q.~Wang, C.~Wang, Q.~Li, C.~Shen, and B.~Feng, ``Veriml: Enabling
  integrity assurances and fair payments for machine learning as a service,''
  \emph{IEEE Transactions on Parallel and Distributed Systems}, vol.~32,
  no.~10, pp. 2524--2540, 2021.

\end{thebibliography}
